\documentclass[10pt,journal]{IEEEtran}

\usepackage[sort]{cite}
\usepackage{color, colortbl, amsmath, amssymb, amsfonts, bm, diagbox}
\usepackage{mathtools}
\usepackage{multirow, graphicx}
\usepackage{algorithm}
\usepackage{algpseudocode}
\usepackage{url, subcaption, stmaryrd, comment, cleveref}
\usepackage{xnewcommand, pifont, url, mathtools, framed}
\usepackage[mmddyyyy]{datetime}
\usepackage[export]{adjustbox}

\graphicspath{{./pic/}}

\def\BibTeX{{\rm B\kern-.05em{\sc i\kern-.025em b}\kern-.08em
    T\kern-.1667em\lower.7ex\hbox{E}\kern-.125emX}}

\newcommand{\sceneone}{S1}
\newcommand{\scenetwo}{S2}
\newcommand{\sceneth}{S3}
\newcommand{\eg}{\mbox{{\em e.g.}}}
\newcommand{\ie}{\mbox{{\em i.e.}}}
\newcommand{\cf}{\mbox{{c.f.}}}
\newcommand{\xmark}{\ding{55}}
\newtheorem{definition}{\bf Definition}

\newcommand[\global]{\deb}[1]{\textcolor{black}{#1}}
\newcommand[\global]{\sv}[1]{\textcolor{black}{#1}}

\settimeformat{ampmtime}

\begin{document}

\title{Feature Engineering for Scalable Application-Level Post-Silicon Debugging}

\author{\IEEEauthorblockN{Debjit Pal\IEEEauthorrefmark{1}, Shobha Vasudevan\IEEEauthorrefmark{2}}
\IEEEauthorblockA{\\
Email: \{\IEEEauthorrefmark{1}dpal2,
\IEEEauthorrefmark{2}shobhav\}@illinois.edu}
}
\markboth{Journal of \LaTeX\ Class Files,~Vol.~14, No.~8, August~2015, Compiled on \today~at \currenttime}%
{Shell \MakeLowercase{\textit{et al.}}: Bare Demo of IEEEtran.cls for IEEE Journals}

\maketitle


\begin{abstract}
We present systematic and efficient solutions for both observability enhancement and root-cause diagnosis of post-silicon System-on-Chips (SoCs) validation with diverse usage scenarios. We model specification of interacting flows in typical applications for message selection. Our method for message selection optimizes {\em flow specification coverage} and {\em trace buffer utilization}. We define the diagnosis problem as identifying buggy traces as {\em outliers} and bug-free traces as {\em inliers/normal behaviors}, for which we use unsupervised learning algorithms for outlier detection. Instead of direct application of machine learning algorithms over trace data using the signals as raw features, we use {\em feature engineering} to transform raw features into more sophisticated features using domain specific operations. The engineered features are {\em highly relevant} to the diagnosis task and are {\em generic} to be applied across any hardware designs. We present debugging and root cause analysis of subtle post-silicon bugs in industry-scale OpenSPARC T2 SoC. We achieve a {\em trace buffer utilization of 98.96\%} with a {\em flow specification coverage of 94.3\%} (average). Our diagnosis method was able to diagnose {\em up to 66.7\%} more bugs and took {\em up to 847$\times$} less diagnosis time as compared to the manual debugging with a diagnosis precision of 0.769.
\end{abstract}

\section{Introduction}

Post-silicon validation is a crucial component of a modern SoC design validation that is performed under highly aggressive schedules and accounts for more than $50\%$ of the validation cost~\cite{cite:yerramili, cite:patra}.

An expensive component of the post-silicon validation is application level {\em use-case validation} through {\em message passing}. In this activity, a validator exercises various target usage scenarios of the system (\eg, for a smartphone, playing videos or surfing the Web, while receiving a phone call) and monitors for failures (\eg, hangs, crashes, deadlocks, overflows, etc.). Each usage scenario involves interleaved execution of several protocols among IPs in the SoC design.
Due to the concurrent execution of multiple protocols~\cite{cite:conf/dac/SingermanAB11, cite:conf/memocode/FraerKKNPSTVY14, cite:conf/dac/AbarbanelSV14}, extremely long execution traces (millions of clock cycles), and lack of bug reproducibility and error sequentiality lead to an extremely time consuming post-silicon diagnosis effort. In current industrial practice~\cite{cite:Mitra:2010:PVO:1837274.1837280}, post-silicon diagnosis is a manual, unsystematic, and ad hoc process primarily relying on the creativity of the validator and often takes weeks to months of validation time. Consequently, it is crucial to determine techniques to streamline this activity.

In this paper, we present an automated post-silicon debug and diagnosis methodology to shorten diagnosis time using machine learning and feature engineering.

In previous work~\cite{DBLP:conf/dac/PalSRPV18}, we developed a {\em message selection method} that specifically targets use-case validation. To debug a use-case scenario, the validator typically needs to observe and comprehend the messages being sent by the constituent IPs. An effective way to do that is to use {\em hardware tracing}~\cite{cite:patra} where a small set of signals are monitored continuously during execution. The effectiveness of hardware tracing is limited by the signals being selected for tracing. Note that the omission of a critical signal manifests only during post-silicon debug when a silicon respin is infeasible.

To select trace signals that are most beneficial for use-case debugging, we depart from the gate-level post-silicon signal selection approaches~\cite{cite:basu2011efficient, cite:chatterjee2011simulation, cite:conf/iccad/MaPJRV15, cite:journals/tvlsi/RahmaniRM17} of prior art and raise the design abstraction at which we apply hardware tracing. We systematically model and analyze usage scenarios at the application level. Our message selection framework uses protocol formalizations as sequences of transactions or {\em flows}~\cite{cite:conf/dac/SingermanAB11, cite:conf/memocode/FraerKKNPSTVY14, cite:conf/dac/AbarbanelSV14, cite:conf/fmcad/TalupurRE15}. Given a collection of usage scenarios and the application-level flows they activate (and the constituent messages), our algorithm computes the messages that are most beneficial for debugging. We scale our observability enhancing algorithms to the industry scale OpenSPARC T2 SoC~\cite{cite:SPARCT2Vol1, cite:SPARCT2Vol2} that is orders of magnitude larger and complex than traditional ISCAS89 benchmarks used in the literature. Along with scale, we argue with empirical evidence that the quality of selected observable are of higher quality and are highly effective for post-silicon usage scenario failure debugging.

Although in~\cite{DBLP:conf/dac/PalSRPV18} we automated the message selection, the debug and diagnosis still remains manual and extremely tedious task. The primary objective of the manual post-silicon debug and diagnosis is i) to understand the desired behavior from the specification, ii) to learn the correct message \deb{patterns} as per the specification, and iii) to learn one or more message \deb{patterns} that are symptomatic of the bug(s). \deb{Machine learning~\cite{DBLP:books/lib/Murphy12, Mitchell:1997:ML:541177} (ML) algorithms automatically learn statistical models from large amounts of training data.}

In this paper, we argue that ML algorithms can learn models of the {\em correct} and {\em buggy} executions of an SoC during the post-silicon debug and diagnosis. To train the models, we can use a large amount of post-silicon trace data that is generated during use-case validation. The primary challenge of applying ML to the diagnosis problem is in representing the training data such that ML model can learn the differences between correct and buggy behavior, and generalize to any arbitrary design.



Logical bugs in designs can be considered as triggering {\em corner-case} design behavior; which is {\em infrequent} and {\em deviant} from normal design behavior. In ML parlance, {\em outlier detection}~\cite{cite:od_survey1, cite:od_survey2} is a technique to identify {\em infrequent} and {\em deviant} data points, called {\em outliers} whereas normal data points are called {\em inliers}. We apply outlier detection techniques to automatically diagnose post-silicon failures by modeling normal design behaviors as inliers and buggy design behavior as outliers.
\sv{Consequently, the task of learning a buggy design behavior transforms into a task of modeling the buggy design behavior as an outlier.}

In post-silicon execution, design bugs typically manifest
as one or more patterns of consecutive messages \deb{(also known as message interleaving)} in the trace data. Human engineers spend considerable time and effort to identify such patterns in the trace data.
We call such a message pattern as an {\em anomalous message sequence}. In this work, we apply ML to identify such anomalous message sequences automatically.

The task for the ML algorithm is the outlier detection where the ML model is expected to learn normal design behaviors and buggy design behaviors and classify buggy behaviors as outliers. \deb{The data used for training is post-silicon execution data where each data point is a triplet consisting of~\textendash~i) the cycle of occurrence of a message, ii) the IP interface from which the message is sourced, and iii) the message iteself} . In our experiments we found that the raw features are insufficient to capture {\em infrequent} and {\em deviant} nature of  buggy design behaviors as compared to the normal behavior (\cf,~\Cref{fig:raw_inability}). 





Feature engineering is a critically important task in the successful generalization of ML to a problem domain~\cite{Heaton_2016,bengio2014representation}. We engineer domain specific features that are highly relevant to the diagnosis task to control the normal and buggy behavior model as seen by the outlier detection algorithms. The engineered features are generic, \ie, they are transformations that can be applied to any hardware designs.

In our formulation we need the anomalous message sequences to appear as outliers. We identify {\em infrequency} and {\em deviancy} as the relevant features that could capture the distinction between normal and anomalous message sequences. Our engineered feature space needs to capture this distinction. We would like normal message sequences to be close and densely distributed in the feature space and anomalous message sequences to be sparsely distributed and distant.






\deb{Due to the large number of possible message sequences, the computation can become computationally expensive. To keep computation tractable, we pre-process trace message sequences to create message aggregates and characterize each such aggregates for anomaly.} A message aggregate with infrequent message sequences contains more information than~\cite{doi:10.1002/j.1538-7305.1948.tb01338.x} a message aggregate with frequent message sequences. We use {\em entropy} to quantify the information content of an aggregate. As the number of infrequent messages sequences in an aggregate increases, the entropy of the aggregate increases monotonically. In order to quantify deviancy of a message sequence with respect to other message sequences in the aggregate, we use a {\em string similarity metric}, in particular {\em Levenshtein distance}~\cite{cite:ldistance}. As an aggregate contains more and more deviant message sequences, the average pairwise Levenshtein distance of the aggregate increases monotonically. We identify message aggregates with both high entropy and high Levenshtein distance as outliers and report them as candidate root causes.

The primary benefits of this diagnosis solution are \textendash~i) it automatically learns the normal and buggy design behaviors from trace message data without training, ii) the engineered features are generic and are independent of any particular design and/or application, and iii) the proposed method can shift through a large amount of trace data, thereby improving detection of candidate anomalous message sequences that are symptomatic of design bugs.

To show scalability and effectiveness of our automated diagnosis approach, we perform our experiments on industrial scale OpenSPARC T2 SoC~\cite{cite:SPARCT2Vol1, cite:SPARCT2Vol2}. We inject complex and subtle bugs, with each bug symptom taking several hundred observed messages ({\em up to 457 messages}) and several hundred thousands of clock cycles ({\em up to 21,290,999 clock cycles}) to manifest. Our analysis shows that the proposed diagnosis method is computationally efficient. It incurred runtime of {\em up to 44.3 seconds} and peak memory usage of {\em up to 508.7 MB} to pre-process trace messages to create aggregates. To detect outlier message aggregates, it incurred runtime of {\em up to 18.91 seconds} and peak memory usage of {\em up to 508.2 MB}.

We also evaluated effectiveness of our engineered features for outlier detection. We found that each of the candidate anomalous message aggregates has entropy of {\em up to 4.3482} and Levenshtein distance of {\em up to 3.0}. This shows that our engineered features are highly effective in demarcating anomalous message aggregates from normal aggregates.

Our analysis shows that our proposed diagnosis method is highly effective. We found that the proposed diagnosis method was able to diagnose {\em up to 66.7\%} more injected bugs with {\em up to 847$\times$} less diagnosis time with a high precision of {\em up to 0.769} as compared to manual debugging.

Our contributions over~\cite{DBLP:conf/dac/PalSRPV18} are as follows. First, we pose the post-silicon diagnosis problem as an outlier detection problem and propose a ML-based scalable and efficient technique to diagnose post-silicon use-case failures.
Second, we systematically model buggy behavior as an outlier and normal behavior as an inlier in the ML data space. To that extent, we engineered two features that are highly relevant to the diagnosis task and applicable across hardware designs.
Third, we establish with empirical evidence that our ML-based technique is highly effective and can diagnose many more bugs at a fraction of time with high precision as compared to manual debugging.

\section{Background and preliminaries}\label{sec:background}

\noindent {\bf Conventions}:  In SoC designs, a message can be viewed as an assignment of Boolean values to the interface signals of a hardware IP. In our formalization below, we leave the definition of message implicit, but we will treat it as a pair $\langle{\mathcal{C}}, w\rangle$ where $w \in \mathbb{Z}^+$.  Informally, ${\mathcal{C}}$ represents the content of the message and $w$ represents the number of bits required to represent ${\mathcal{C}}$.  Given a message $m=\langle {\mathcal{C}},w\rangle$, we will refer to $w$ as {\em bit-width of $m$}, denoted by $\mbox{{\em width}}(m)$ or $|m|$.


\begin{figure}
	\centering
	\begin{subfigure}[b]{0.15\textwidth}
		\includegraphics[scale=0.35]{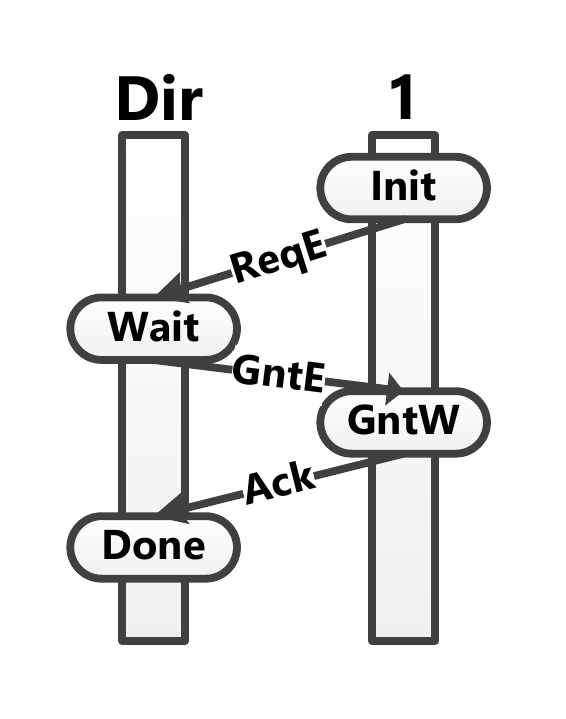}
		\caption{\label{fig:flow}}
	\end{subfigure}
	\hfill
	\begin{subfigure}[b]{0.3\textwidth}
		\includegraphics[scale=0.3]{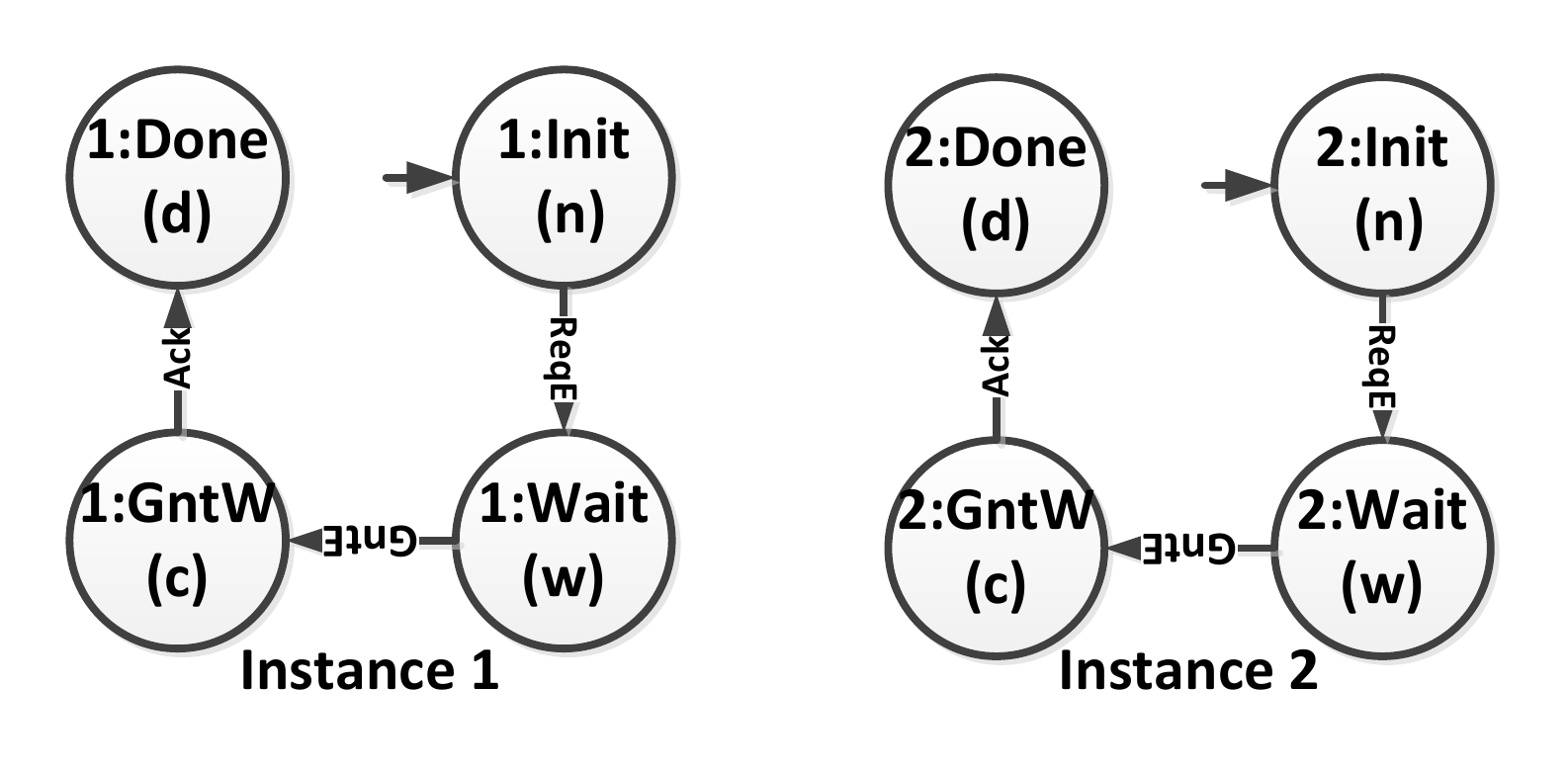}
		\caption{\label{fig:tagged}}
	\end{subfigure}
	\caption{\ref{fig:flow} shows a {\em flow} for an exclusive line access request for a cache coherence flow~\cite{cite:conf/fmcad/TalupurRE15} along with participating IPs.~\ref{fig:tagged} shows two legally indexed instances of cache coherence flow.}
	\vspace{-2mm}
\end{figure}

\begin{definition}\label{def:flow}
 A {\bf flow} is a directed acyclic graph (DAG) defined as a tuple, $\mathcal{F} = \langle \mathcal{S}, \mathcal{S}_0, \mathcal{S}_{p}, \mathcal{E}, \delta_{\mathcal{F}}, Atom \rangle$ where $\mathcal{S}$ is the set of flow states, $\mathcal{S}_0 \subseteq \mathcal{S}$ is the set of initial states, $\mathcal{S}_p \subseteq \mathcal{S}~\text{and}~\mathcal{S}_p \cap Atom = \emptyset$~is called the set of stop states, $\mathcal{E}$ is a set of messages,  $\delta_\mathcal{F} \subseteq \mathcal{S} \times \mathcal{E} \times \mathcal{S}$ is the transition relation and $Atom \subset \mathcal{S}$ is the set of atomic states of the flow.
\end{definition}

\smallskip

We use $\mathcal{F}.\mathcal{S}, \mathcal{F}.\mathcal{E}$ etc. to denote the individual components of a flow $\mathcal{F}$. A {\em stop} state of a flow is its final state after its successful completion. $Atom$ is a {\em mutex} set of flow states \ie, any two flow states in $Atom$ cannot happen together. Other components of $\mathcal{F}$ are self-explanatory. In Figure~\ref{fig:flow}, we have shown a toy cache coherence flow along with the participating IPs and the messages. In Figure~\ref{fig:flow},
$\mathcal{S} = $ \{Init, Wait, GntW, Done\}, $\mathcal{S}_0 = $ \{Init\}, $\mathcal{S}_p$ = \{Done\}, $Atom = $ \{GntW\}. Each of the messages in the cache coherence flow is 1 bit wide, hence $\mathcal{E} = \{\langle ReqE, 1 \rangle, \langle GntE, 1 \rangle, \langle Ack, 1 \rangle\}$.

\medskip


\begin{definition}\label{def:exe_flow}
 Given a flow $\mathcal{F}$, an {\bf execution} $\rho$ is an alternating sequence of flow states and messages ending with a stop state. For flow $\mathcal{F}$, $\rho = s_0~\alpha_1~s_1~\alpha_2~s_2~\alpha_3~\ldots~\alpha_n~s_n$ such that $s_i \overset{\alpha_{i + 1}}{\longrightarrow} s_{i + 1}, \forall 0 \leq i < n,~s_i \in \mathcal{F}.S, ~\alpha_{i + 1} \in \mathcal{F}.\mathcal{E},~s_n \in \mathcal{F}.\mathcal{S}_p$. The {\bf trace} of an execution $\rho$ is defined as $trace(\rho) = \alpha_1~\alpha_2~\alpha_3~\alpha_4~\ldots~\alpha_n$.
\end{definition}

\smallskip

%
%
%

\begin{figure}
	\centering
	\includegraphics[scale=0.45]{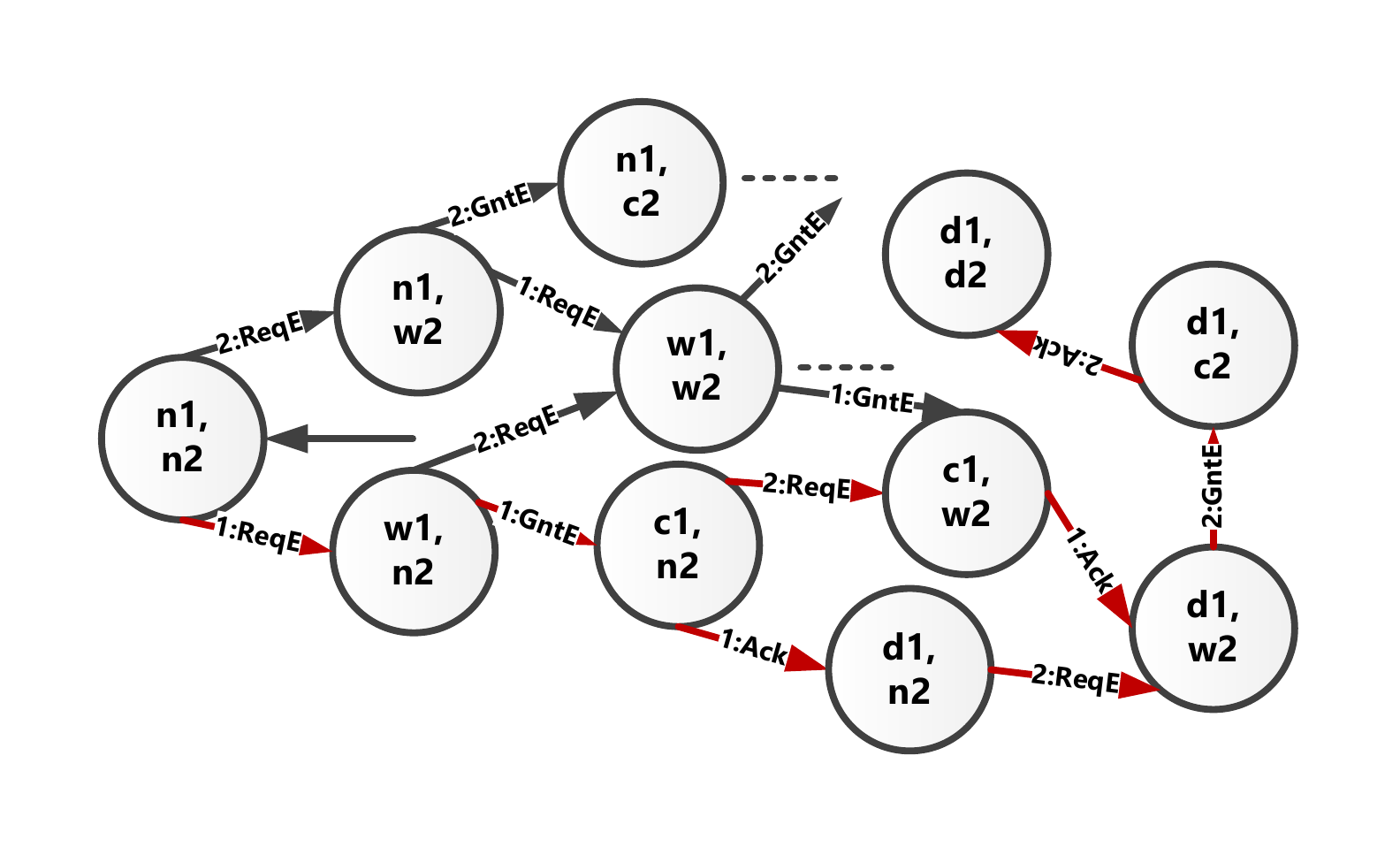}
	\caption{Two instances of cache coherence flow of Figure~\ref{fig:flow}
     interleaved.\label{fig:intra_interleaved}}
     \vspace{-1mm}
\end{figure}

For the cache coherence flow of Figure~\ref{fig:flow}, $\rho$ = \{{\em n, ReqE, w, GntE, c, Ack, d}\}, $trace(\rho) = $ \{{\em ReqE, GntE, Ack}\}.


Intuitively, a flow provides a pattern of system execution. A flow can be invoked several times, even concurrently, during a single run of the system. To make precise the relation between an execution of the system with participating flows, we need to distinguish between these instances of the same flow. The notion of {\em indexing} accomplishes that by augmenting a flow with an ``index".

\medskip

\begin{definition}\label{def:idx_msg_flow}
 An {\bf indexed message} is a pair $\alpha = \langle m, i \rangle$ where $m$ is the message and $i \in \mathbb{N}$, referred to as the {\bf index} of $\alpha$. An {\bf indexed state} is a pair $\hat{s} = \langle s, j \rangle$ where $s$ is a flow state and $j \in \mathbb{N}$, referred as the index of $\hat{s}$. An {\bf indexed flow} $\langle f, k \rangle$ is a flow consisting of indexed message $m$ and indexed state $\hat{s}$ indexed by $k \in \mathbb{N}$.
\end{definition}

\smallskip

Figure~\ref{fig:tagged} shows two instances of the cache coherence flow of Figure~\ref{fig:flow} indexed with their respective instance number. In our modeling, we ensure by construction that two different instances of the same flow do not have same indices. Note that in practice, most SoC designs include architectural  support to enable {\em tagging}, \ie, uniquely identifying different concurrently executing instances of the same flow. Our formalization simply makes the notion of tagging explicit.

\medskip

\begin{definition}\label{def:legal_idx}
 Any two indexed flows $\langle \mathcal{F}, i \rangle$, $\langle \mathcal{G}, j \rangle$ are said to be {\bf legally indexed} either if $\mathcal{F} \neq \mathcal{G}$ or if $\mathcal{F} = \mathcal{G}$ then $i \neq j$.
\end{definition}

\smallskip

Figure~\ref{fig:tagged} shows two legally indexed instances of the cache coherence flow of
Figure~\ref{fig:flow}. Indices uniquely identify each instance of the cache coherence flow.

\medskip

A {\bf usage scenario} is a pattern of frequently used applications. Each such pattern comprises multiple interleaved flows corresponding to communicating hardware IPs.

\smallskip

\begin{definition}\label{def:intra_interleaved}
Let $\mathcal{F}, \mathcal{G}$ be two legally indexed flows. The interleaving $\mathcal{F} \interleave \mathcal{G}$ is a flow called {\bf interleaved flow} defined as~$\mathcal{U} = \mathcal{F} \interleave \mathcal{G} = \langle \mathcal{F}.\mathcal{S} \times \mathcal{G}.\mathcal{S}, \mathcal{F}.\mathcal{S}_0 \times \mathcal{G}.\mathcal{S}_0, \mathcal{F}.\mathcal{S}_p \times \mathcal{G}.\mathcal{S}_p, \mathcal{F}.\mathcal{E} \cup \mathcal{G}.\mathcal{E}, \delta_{\mathcal{U}}, \mathcal{F}.Atom \cup \mathcal{G}.Atom \rangle$ where $\delta_{\mathcal{U}}$ is defined as:
	\begin{center}
		i) $\frac{s_1 \overset{\alpha}{\longrightarrow} s_1^\prime~\wedge~s_2 \not\in \mathcal{G}.Atom}{\langle s_1,
        s_2 \rangle \overset{\alpha}{\longrightarrow} \langle s_1^\prime, s_2 \rangle}$ \quad and \quad ii) $\frac{s_2 \overset{\beta}{\longrightarrow} s_2^\prime~\wedge~s_1 \not\in \mathcal{F}.Atom}{\langle s_1, s_2 \rangle	\overset{\beta}{\longrightarrow} \langle s_1, s_2^\prime \rangle}$
	\end{center}
	where $s_1, s_1^\prime \in \mathcal{F}.\mathcal{S}$, $s_2, s_2^\prime \in \mathcal{G}.\mathcal{S}$, $\alpha \in \mathcal{F}.\mathcal{E}$, $\beta \in \mathcal{G}.\mathcal{E}$. Every path in the interleaved flow is an execution of $\mathcal{U}$ and represents an interleaving of the messages of the participating flows.
\end{definition}

\smallskip

Rule i) of $\delta_{\mathcal{U}}$ says that if $s_1$ evolves to the state $s_1^\prime$ when message $\alpha$ is performed and if $g$ has a state $s_2$ which is not atomic/indivisible, then in the interleaved flow, if we have a state  $(s_1, s_2)$, it evolves to state $(s_1^\prime, s_2)$ when message $\alpha$ is performed. A similar explanation holds good for Rule ii) of $\delta_{\mathcal{U}}$. For any two concurrently executing legally indexed flow $\mathcal{F}$ and $\mathcal{G}$, $J = \mathcal{F} \interleave \mathcal{G}$, for any $s \in \mathcal{F}.Atom$ and for any $s^\prime \in \mathcal{G}.Atom$, $(s, s^\prime) \not\in J.\mathcal{S}$. If one flow is in one of its atomic/indivisible state, then no other concurrently executing flow can be in its atomic/indivisible state.

Figure~\ref{fig:intra_interleaved} shows partial interleaving $\mathcal{U}$ of two legally indexed flow instances of Figure~\ref{fig:tagged}. Since $c_1$ and $c_2$  both are atomic state, state $(c_1, c_2)$ is an illegal state in the interleaved flow. $\delta_{\mathcal{U}}$ and the $Atom$ set make sure that such illegal states do not appear in the interleaved flows.

Trace buffer availability is measured in terms of bits thus rendering bit width of a message important. In Definition~\ref{def:msg_comb}, we define a message combination. Different instances of the same message i.e. indexed messages are not required while computing the bit width of the message combination.

\medskip

\begin{definition}\label{def:msg_comb}
A {\bf message combination} $\mathcal{M}$ is an unordered set of messages. The {\bf total bit width} $W$ of a message combination $\mathcal{M}$ is the sum total of the bit width of the individual messages contained in $\mathcal{M}$ i.e. $W(\mathcal{M}) = \sum_{i = 1}^{k} width(m_i) = \sum_{i = 1}^{k} w_i, m_i \in \mathcal{M}, k = |\mathcal{M}|$.
\end{definition}

\smallskip

We introduce a metric called {\bf flow specification coverage} to evaluate the quality of a {\em message combination}.

\medskip

\begin{definition}\label{def:flow_spec_cov}
Let $\mathcal{F}$ be a flow. The {\bf visible flow states} $visible(\alpha)$ of a message $\alpha \in \mathcal{F}.\mathcal{E}$ is defined as the set of flow states reached on the occurrence of message $\alpha$ \ie, $visible(\alpha) = \{s^\prime | s \overset{\alpha}{\longrightarrow} s^\prime, s, s^\prime \in \mathcal{F}.\mathcal{S}\}$. The {\bf flow specification coverage} $FCov(\mathcal{M})$ of a message combination $\mathcal{M}$ is defined as the set union of the visible flow states of all the messages in the message combination, expressed as a fraction of the total number of flow states \ie, $FCov(\mathcal{M}) = \frac{\displaystyle\cup_{i=1}^k visible(\alpha_i)}{|\mathcal{F}.\mathcal{S}|}, k = |\mathcal{M}|$.
\end{definition}

\smallskip

We extend the definition of a {\em trace}($\rho$) of an {\em execution} $\rho$ (\cf,~\Cref{def:exe_flow}) to define {\em message sequence} and {\em message aggregate} for diagnosis.

\medskip

\begin{definition}\label{def:m_sequence}
 A {\bf message sequence} $m(\rho)$ of a  trace($\rho$) is defined as a subsequence of the trace of the execution. The {\bf length} $k$ of a message sequence $m(\rho)$ is defined as the number of messages contained in $m(\rho)$. For example, for $trace(\rho) = \alpha_1~\alpha_2~\alpha_3~\ldots~\alpha_n$, $m(\rho) = \langle \alpha_1~\alpha_2~\alpha_3 \rangle$ is a message sequence of trace($\rho$) of length $k = 3$. Any two message sequences $m_i(\rho)$ and $m_j(\rho)$ of length $k$ are {\bf distinct} if $\exists l \in [1, k], \alpha_{i, l} \neq \alpha_{j, l}$ where $\alpha_{i, l } \in m_i(\rho)$, $\alpha_{j, l } \in m_j(\rho)$.
\end{definition}

\smallskip

\begin{definition}\label{def:m_aggregate}
 A {\bf message aggregate} $maggr(\rho)$ of a trace($\rho$) is defined as an unordered set of message sequences of length $k$. Each distinct message sequence in a message aggregate is called an {\bf unique message sequence} of that message aggregate. For example,  $maggr(\rho) = \{\langle \alpha_1~\alpha_2~\alpha_3 \rangle, \langle \alpha_2~\alpha_3~\alpha_4 \rangle\}$ is a message aggregate of length 3 message sequences of trace($\rho$). Each of the $\langle \alpha_1~\alpha_2~\alpha_3 \rangle$ and $\langle \alpha_2~\alpha_3~\alpha_4 \rangle$ is an unique message sequence of $maggr(\rho)$.
\end{definition}

For the cache coherence flow of~\Cref{fig:flow}, $m_1(\rho)$ = $\langle ReqE,GntW \rangle$, $m_2(\rho)$ = $\langle GntW,Ack \rangle$ are two length-2 message sequences and $maggr(\rho)$ = $\{m_1,m_2\}$ = $\{\langle ReqE, GntW \rangle,\langle GntW, Ack \rangle\}$ is a message aggregate. 

\subsection{Entropy and mutual information gain}\label{sec:entropyinfogain}

\noindent{\bf Entropy}: The {\em entropy}~\cite{doi:10.1002/j.1538-7305.1948.tb01338.x} measures the uncertainty in a random variable. Let $X$ be a discrete random variable with possible values $X_{val} = \{x_1, x_2, \ldots, x_n \}$. Let $p(x)$ be the associated probability mass function of $X$. The entropy of the random variable $X$ is defined as $\mathcal{H}(X) = -\textstyle\sum_{x_i \in X_{val}}p(x_i)log_2 p(x_i)$ where $p(x_i) = |X = x_i|/|X_{val}|$ denotes the fraction of $X$ in which $X = x_i$.

\smallskip

\noindent{\bf Mutual information gain}: The {\em Mutual information gain}~\cite{doi:10.1002/j.1538-7305.1948.tb01338.x} measures the amount of information that can be obtained about one random variable $X$ by observing another random variable $Y$. More precisely, the conditional entropy of a random variable $X$ with respect to another random variable $Y$ is the reduction in uncertainty in the realization of $X$ when the outcome of $Y$ is known. For jointly distributed discrete random variables $X$ and $Y$, the mutual information gain of $X$ relative to $Y$ is given by $I(X; Y) = \textstyle\sum_{x, y} p(x, y)log_2(p(x, y)/(p(x)p(y)))$, where $p(x)$ and $p(y)$ are the associated random probability mass function for two random variables X and Y respectively.


\subsection{Levenshtein distance}\label{sec:ldistance}

The Levenshtein distance is a {\em string similarity metric} for measuring the {\em dissimilarity} between two strings.
Mathematically, the Levenshtein distance between two strings $a, b$ (of length $|a|$ and $|b|$) $\mathcal{L}_{a, b}(|a|, |b|)$ is defined as:
\[  \scriptsize
    \mathcal{L}_{a, b}(i, j) = \begin{cases}
                            max(i, j) & \mbox{if } min(i, j) = 0 \\
                            min \begin{cases}
                                 \mathcal{L}_{a, b}(i - 1, j) &\\
                                 \mathcal{L}_{a, b}(i, j - 1) &\\
                                 \mathcal{L}_{a, b}(i - 1, j - 1) + \mathbf{1}_{(a_i \neq b_j)} &
                                \end{cases} & \mbox{otherwise}.
                           \end{cases}
\]

where $\mathbf{1}_{(a_i \neq b_j)}$ is the {\em indicator function} equal to $0$ when $a_i == b_j$ and equal to $1$ otherwise. The $\mathcal{L}_{a, b}(i, j)$ is the distance between the first $i$ characters of $a$ and the first $j$ characters of $b$.
We will denote $\mathcal{L}_{a, b}(|a|, |b|)$ as $\mathcal{L}(a, b)$.

\smallskip



The salient features of Levenshtein distance are \textendash~i) it is at least the difference of the sizes of the two strings, ii) it is at most the length of the longer string, iii) it is zero {\em iff} the strings are equal, and iv) if the strings are of the same size, the Hamming distance~\cite{cite:hamming} is an upper bound.

\section{Outlier detection}\label{sec:anomalyalgo}

\subsection{Outliers in ML}\label{sec:what_is_outlier}
In ML, outliers are defined as data samples whose characteristics notably deviate from our expectation \cite{cite:od_survey1, cite:od_survey2}. Outliers have two characteristics \textendash~i) they are different and
ii) they are rare as compared to normal data samples.
%
%

In spite of straightforward definition, detecting an outlier is challenging. First, the boundary between outliers and normal samples are often not precise. Additionally, some outliers only manifest their outlierness in an engineered feature space derived from the original feature space via non-trivial transformation. Second, the groundtruth of the outliers is often absent due to prohibitive cost. Hence in many cases, outliers are determined based on the sample characteristics alone. Unsupervised outlier detection algorithms are developed to identify outliers through only the patterns and intrinsic properties of the feature space, and hence do not require any groundtruth labels.
%
%
%
\subsection{Different notions of outliers}\label{sec:conceptual_outlier}
There are three distinct notions of outliers depending on the profiling of normal samples~\textendash i) {\em classification-based}, ii) {\em density-based}, and iii) {\em spectral-based}.
\smallskip

\noindent {\bf Classification-based outlier detection}: Outliers can be defined by a classifier which can be learned in the feature space to distinguish between the normal and anomalous samples \cite{cite:od_survey2}. Any sample that does not fit the representation of the normal samples  would be considered as outliers. When the groundtruth is unavailable, the classifier can be learned in an unsupervised manner. \textit{One-class Support Vector Machine} (OCSVM) \cite{cite:ocsvm, cite:ocsvm2} is an unsupervised outlier detection method that adopts this notion of outliers.

\smallskip

\noindent {\bf Density-based outlier detection}: Density-based outliers are based on the assumption that the normal data samples reside in neighborhoods of high density whereas outliers reside in low-density regions \cite{cite:od_survey2}. There are two distinct notion of density-based outlierness. First, the local density of a sample can be estimated as its distance to its $k$-nearest neighbors,\footnote{The distance can be the distance to the $k^{th}$ distant neighbor or the average of distances of all {\em k} neighbors.} with larger distances indicating higher degrees of outlierness. The {\em k-Nearest Neighbors} ($k$NN) \cite{cite:knn1, cite:knn2} is an unsupervised outlier detection technique that adopts this notion of outliers directly and uses the distance to quantify outlierness. Second, the relative density of each data sample to the density of its $k$ neighbors can be used as an indication of outlierness of a sample. A normal sample has a local density that is similar to its neighbors whereas an outlier's local density is lower than that of its neighbors. \textit{Local Outlier Factor} (LOF) \cite{cite:lof} is an unsupervised outlier detection method that identifies outliers based on the relative density of a sample's neighborhoods.

\smallskip

\noindent {\bf Spectral-based outlier detection}: Spectral-based notion of outliers assumes that the difference between normal samples and outliers can be significantly enhanced when the data is embedded into a lower dimensional subspace \cite{cite:od_survey2}. Hence, outlier detection methods that adopt this notion of outliers approximate the data space using a transformation of the original features to capture the variability in the data for easy outlier identification. \textit{Principal Component Analysis} (PCA) \cite{cite:pca} is an unsupervised outlier detection method that can project data into a lower dimensional space where most of the variability of the data is captured and explained by the new dimensions. The variability that is not captured by the new dimensions is considered as anomalous. {\em Isolation Forest} (IForest) \cite{cite:iforest1, cite:iforest2} is another unsupervised outlier detection method that attempts to identify outliers using only a subset of the features. IForest recursively select feature and split feature values in random until samples are isolated. Since outliers are rare and lie further away from the normal samples in the feature space, the number of splittings required to isolate an outlier can be served as its outlier score.

\subsection{Metrics of an outlier detection algorithm}\label{sec:precision}

\begin{definition}\label{def:precision}
  The precision of an outlier detection algorithm is defined as the number of true positives expressed as a fraction of total number of samples labeled as belonging to the outlier class \ie, $Precision = \frac{t_p}{t_p + f_p}$ where $t_p$ = number of true positives, $f_p$ = number of false positives.AP
\end{definition}

\smallskip

\begin{definition}\label{def:recall}
  The recall of an outlier detection algorithm is defined as the number of true positives expressed as a fraction of total number of true positives and false negatives \ie, $Recall = \frac{t_p}{t_p + f_n}$, where $f_n$ = number of false negatives.
\end{definition}

\smallskip

\begin{definition}\label{def:accuracy}
  The accuracy of an outlier detection algorithm is defined as the number of samples that are correctly labeled as belonging to both the outlier class and normal class expressed as a fraction of total number of samples \ie, $Accuracy = \frac{t_p + t_n}{t_p + t_n + f_p + f_n}$ , where $t_n$ = number of true negatives.
\end{definition}

\section{Message selection methodology}\label{sec:MsgSel}

\subsection{Objective of message selection methodology}\label{sec:prob_statement}

Maximizing information gain is done in order to increase flow specification coverage during post-silicon debug of usage scenarios. The message selection procedure considers the {\em message combination} $\mathcal{M}$ for tracing, whereas to calculate information gain over $\mathcal{U}$, it uses {\em indexed messages}.

Given a set of legally indexed participating flows of a usage scenario, bit widths of associated messages, and a trace buffer width constraint, {\bf our method selects a message combination such that information gain is maximized over the interleaved flow} $\mathcal{U}$ and the {\bf trace buffer is maximally utilized}.

For the cache coherence flow example of Figure~\ref{fig:flow}, we assume a trace buffer width of 2 bits and concurrent execution of two instances of the flow. {\em ReqE, GntE}, and {\em Ack} messages happen between {\em 1-Dir, Dir-1}, and {\em 1-Dir} IP pairs respectively. {\em ReqE, GntE}, and {\em Ack} consist of {\tt req, gnt} and {\tt ack} signal and each of the messages is 1-bit wide. Let $\mathbb{B} = \{0,1\}$ be the set of Boolean values. $\mathcal{C}(ReqE) = \mathbb{B}^{|req|}$, $\mathcal{C}(GntE) = \mathbb{B}^{|gnt|}$, and $\mathcal{C}(Ack) = \mathbb{B}^{|ack|}$ denote respective message contents.

\begin{figure}
  \centering
  \includegraphics[scale=0.6]{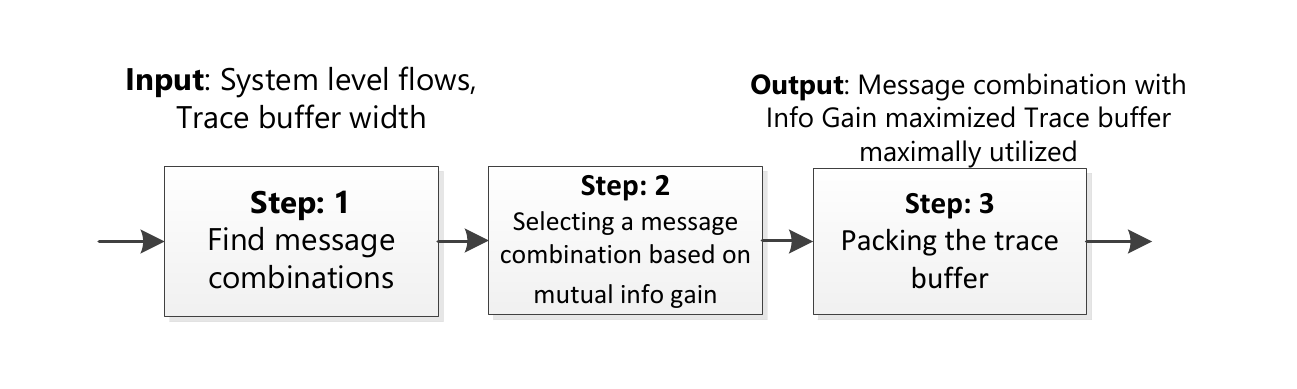}
  \caption{Our message selection approach\label{fig:MsgSelIterSplit}}
  \vspace{-2mm}
\end{figure}

\subsection{{\bf Step 1}: Finding message combinations}\label{subsec:findMsgComb}

In Step 1, we identify all possible message combinations from the set of all messages of the participating flows in a usage scenario.

While we find different message combinations, we also calculate the total bit width of each such combination. Any message combination that has a total bit width less than or equal to the available trace buffer width is kept for further analysis in Step 2.\footnote{For multi-cycle messages, the number of bits that can be traced in a single cycle is considered as the message bit width} Each such message combination is a potential candidate for tracing.

In the example of Figure~\ref{fig:flow}, there are 3 messages and $\sum_{k=1}^{3} \binom 3 k = 7$ different message combinations. Of these, only one ({\em ReqE, GntE, Ack}) has a bit width more than trace buffer width (2). We retain the remaining six message combinations for further analysis in Step 2.
\begin{figure*}
    \begin{subfigure}[b]{0.30\textwidth}
        \centering
        \includegraphics[scale=0.35]{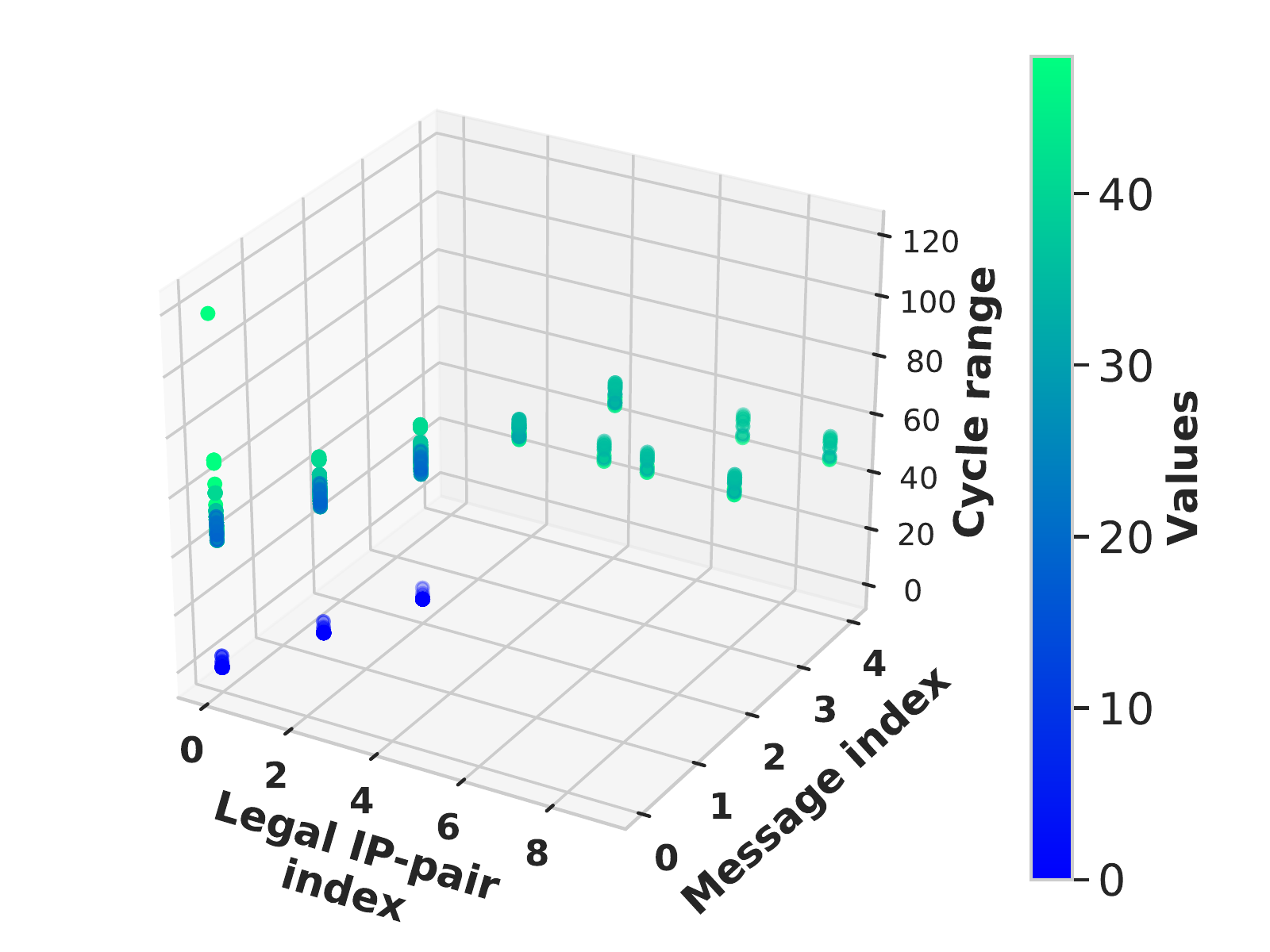}
        \caption{Case study 1 ($k = 5$)\label{fig:test1_raw_scatter}}
    \end{subfigure}
    \hfill
    \begin{subfigure}[b]{0.30\textwidth}
        \centering
        \includegraphics[scale=0.35]{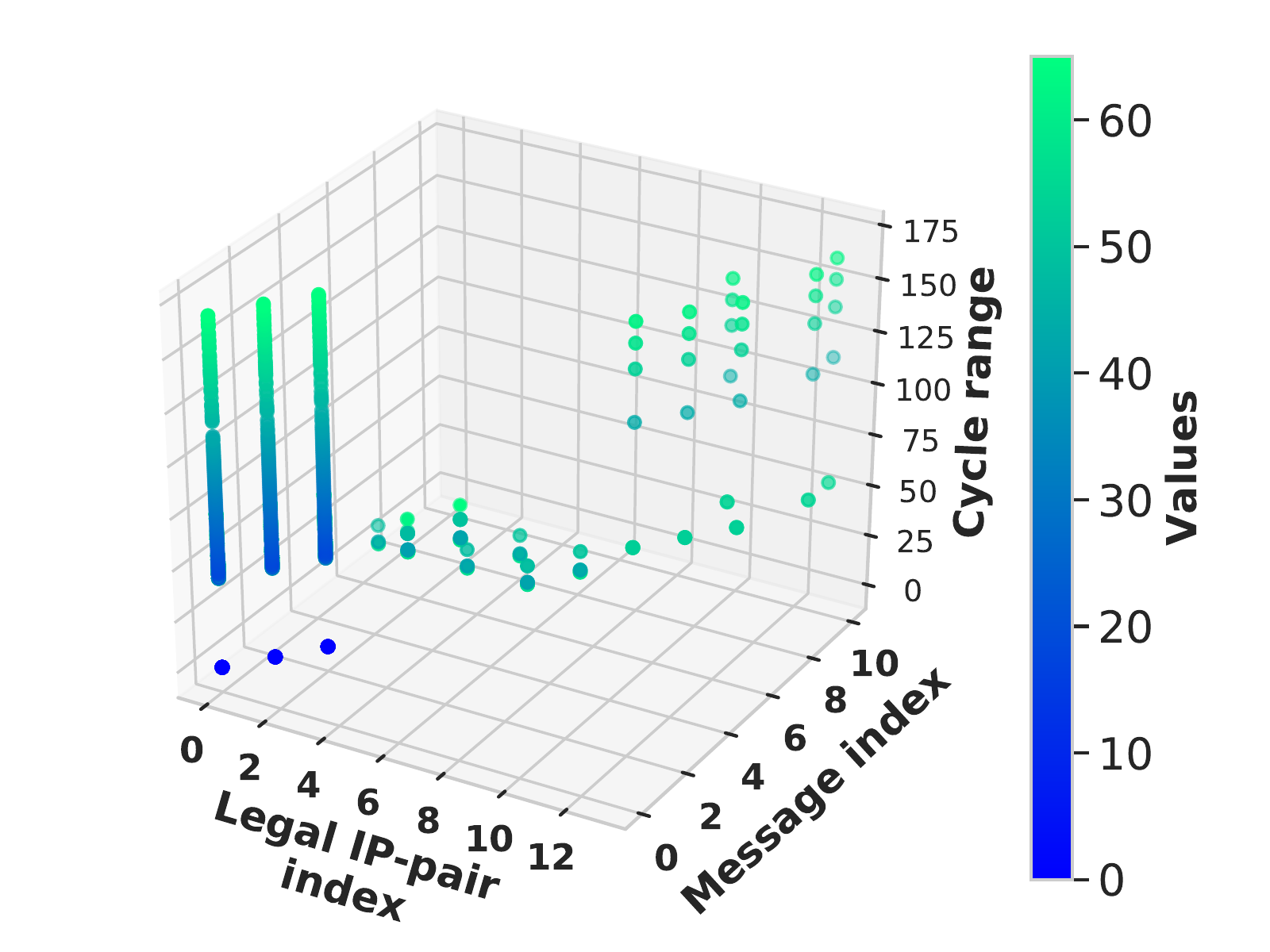}
        \caption{Case study 3 ($k = 5$)\label{fig:test3_raw_scatter}}
    \end{subfigure}
    \hfill
    \begin{subfigure}[b]{0.30\textwidth}
        \centering
        \includegraphics[scale=0.35]{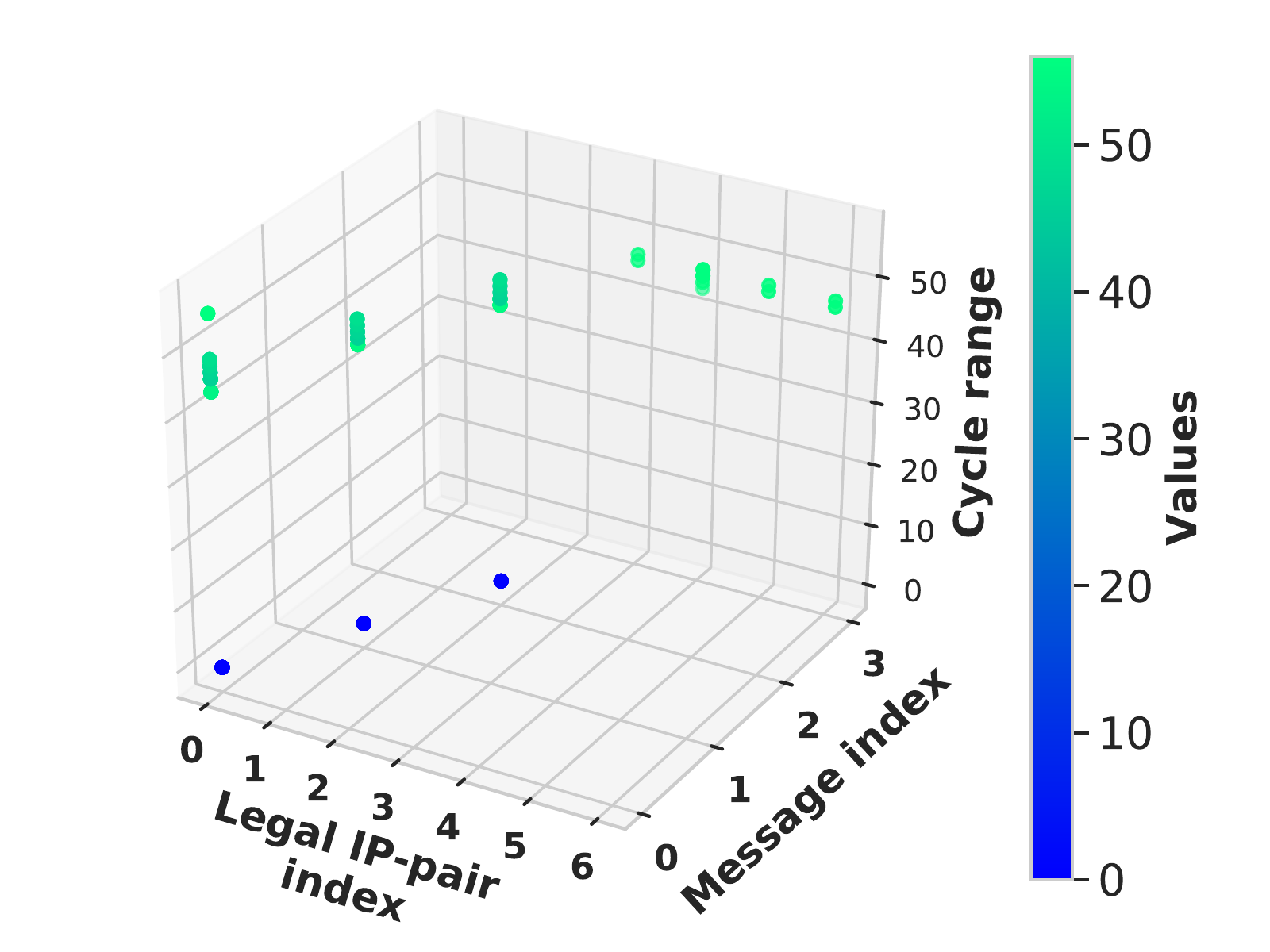}
        \caption{Case study 5 ($k = 5$)\label{fig:test5_raw_scatter}}
    \end{subfigure}
    \caption{(\subref{fig:test1_raw_scatter}),~(\subref{fig:test3_raw_scatter}), and~(\subref{fig:test5_raw_scatter}) show inability of raw feature data to demarcate anomalous message sequences.\label{fig:raw_inability}}
    \vspace{-6mm}
\end{figure*}

\subsection{{\bf Step 2}: Selecting a message combination based on mutual information gain}\label{sec:mutualinfogain}

In this step, we compute the mutual information gain of message combinations computed in step 1 over the interleaved flow. We then select the message combination that has the {\bf highest mutual information gain} for tracing.

We use {\em mutual information gain} as a metric to evaluate the quality of the selected set of messages with respect to the interleaving of a set of flows. We associate two random variables with the interleaved flow namely $X$ and $Y_i$. $X$ represents the different states in the interleaved flow i.e. it can take any value in the set $\mathcal{S}$ of the different states of the interleaved flow. Let $\mathcal{M}  = \bigcup_{i} \mathcal{E}_i$ be the set of all possible indexed messages in the interleaved flow. Let $Y_i^\prime$ be a
candidate message combination and $Y_i$ be a random variable representing all indexed messages corresponding to $Y_i^\prime$. All values of $X$ are equally probable since the interleaved flow can be in any state and hence $p_X(x) = \frac{1}{|\mathcal{S}|}$. To find the marginal distribution of $Y_i$, we count the number of occurrences of each indexed message in the set $\mathcal{M}^\prime$ over the entire interleaved flow. We define $p_{Y_i}(y) = \frac{\text{\# of occurrences of y in flow}}{\text{\# of occurrences of all indexed messages in flow}}$. To find the joint probability, we use the conditional probability and the marginal distribution i.e. $p(x,y) = p(x|y)p(y) = p(y|x)p(x)$. $P(x|y)$ can be calculated as the fraction of the interleaved flow states $x$ is reached after the message $Y_i = y$ has been observed. In other words, $p(x|y)$ is the fraction of times $x$ is reached, from the total number of occurrences of the indexed message $y$ in the interleaved flow i.e. $p_{X|Y_i}(x|y) = \frac{\text{\# occurrence of y in flow leading to x}}{\text{total \# occurrences of y in flow}}$. Now we substitute these values in $I(X;Y)$ to calculate the mutual information gain of the state set $X$ w.r.t $Y_i$.

In Figure~\ref{fig:intra_interleaved}, $p_X(x) = \frac{1}{15} \forall x \in \mathcal{S}$. Let $Y_1^\prime = \{GntE, ReqE\}$ be a candidate message combination and $Y_1$ = \{{\em 1:GntE, 2:GntE, 1:ReqE, 2:ReqE}\}. For $I(X;Y_1)$, we have $p(y = y_i) = \frac{3}{18}, \forall y_i \in Y_1$. Therefore, $p_{X|Y_1}(x|1:GntE) = \{1/3~if~x = (c1, n2), 1/3~if~x = (c1, w2), 1/3~if~x = (c1, d2)\}$ and $p_{X,Y_1}(x, 1:GntE) = \{1/18~if~x = (c1, n2), 1/18~if~x = (c1, w2), 1/18~if~x = (c1, d2)\}$. Similarly, we calculate $p_{X, Y_1}(x, 2:GntE)$, $p_{X, Y_1}(x, 1:ReqE)$ and $p_{X, Y_1}(x, 2:ReqE)$. The mutual information gain is given by: $I(X, Y_1) = \sum_{x,y}p(x,y) log p(x,y)/p(x)p(y) = 1.073$.

Similarly, we calculate the mutual information gain for the remaining five message combinations. We then select the message combination that has the highest mutual information gain, which is $I(X, Y_1) = 1.073$ thereby selecting the message combination $Y_1^\prime = \{\text{ReqE, GntE}\}$ for tracing. Intuitively, in an execution of $\mathcal{U}$ of Figure~\ref{fig:intra_interleaved}, if the observed trace is \{{\em 1:ReqE, 1:GntE, 2:ReqE}\}, immediately we are able to localize the execution to two paths shown in red in
Figure~\ref{fig:intra_interleaved} among many possible paths of $\mathcal{U}$.

\subsection{{\bf Step 3}: Packing the trace buffer}\label{subsec:maxutilization}

Message combinations with the highest mutual information gain selected in Step 2 may not completely fill the trace buffer.  To maximize trace buffer utilization, in this step we {\it pack} smaller message groups which are small enough to fit in the leftover trace buffer width. Usually, these smaller message groups are part of a larger message that cannot be fit into the trace buffer, \eg~in OpenSPARC T2, {\tt dmusiidata} is 20 bits wide message whereas {\tt cputhreadid} a subgroup of {\tt dmusiidata} is 6 bits wide. We select a message
group that can fit into the leftover trace buffer width, such that the information gain of the selected message combination in union with this smaller message group is maximal. We repeat this step until no more smaller message groups can be added in the leftover trace buffer. Benefits of packing are shown empirically in Section ~\ref{sec:funcspeccov}.

In our running example, the trace buffer is filled up by the set of selected message combination. The flow specification coverage achieved with $Y_1^\prime$ is 0.7333.


\section{Bug symptom diagnosis methodology}\label{sec:detectad}

\subsection{Formulation of post-silicon debug as an outlier detection problem}\label{sec:psi_given}
A post-silicon execution {\em passes} if it finishes without any failures \eg, hangs, deadlock, livelock, crash etc., otherwise the execution {\em fails}. For the diagnosis problem, we consider traced messages during execution as input data. In post-silicon execution, a failure happens due to the occurrence of one or more message sequence(s) that is symptomatic of one or more design bugs. We consider such a message sequence as an {\em anomalous message sequence}. Since an anomalous message sequence represents a deviant design behavior, we consider such a message sequence as an {\em outlier} in post-silicon execution data space. Consequently, we formulate post-silicon diagnosis as an outlier detection problem. {\bf Given a set of anomalous post-silicon executions, our diagnosis method identifies one or more candidate anomalous message sequences.}

Since post-silicon executions span millions of clock cycles, hence for tractable computation, we segregate raw trace data in multiple cycle ranges. Further, we assign an index to every {\em legal} IP pair\footnote{An IP pair is {\em legal} if a message is passed between them.} and to every unique message that happens in a post-silicon execution.\footnote{This index is an enumeration of traced messages and is different from indexed messages discussed in~\Cref{def:idx_msg_flow}.} The segregated trace data has three raw features \textendash~i) cycle range in which the message has occurred, ii) the index of the legal IP-pair between which the message has occurred, and iii) the index of a message that has occurred. In~\Cref{fig:raw_inability} we show raw trace data in three-dimensional feature space for several case studies (\cf,~\Cref{sec:exp_results_detection}) for OpenSPARC T2 SoC.

\subsection{Insufficiency of raw features for detection}\label{sec:bugs_as_anomalies}
An anomalous message sequence has two primary characteristics \textendash~i) it is {\em infrequent} and ii) it is {\em deviant} from other normal message sequences. An in-depth inspection of~\Cref{fig:raw_inability} shows that the trace data in raw feature space has the following deficiencies \textendash~i) the raw features provide message-specific information, ii) in raw feature space outliers are not well demarcated, and iii) the raw features fail to provide context of the failure during diagnosis.

Hence, we pre-process raw trace message data to construct message sequences and characterize each such message sequences for {\em infrequency} and {\em deviancy} using engineered features (\cf,~\Cref{sec:what_is_outlier}). \deb{The computational cost of analyzing each of the individual message sequences can be prohibitively large due to the large number of message sequences obtained from traces.} To \deb{keep computational cost nominal}, instead of analyzing each of the message sequences individually, we analyze message aggregates of message sequences and characterize each such aggregate for the anomaly.

\subsection{Intuition of engineered features}\label{sec:engg_feature_intuition}

In order to quantify the characterization of anomalousness, we calculate two engineered feature values of each of the message aggregates \textendash~i) entropy (characterizes infrequency) and ii) Levenshtein distance (characterizes deviancy).

\medskip

\noindent{\bf Entropy as an engineered feature}: A message aggregate is characterized as anomalous if it contains one or more infrequent unique message sequences. An aggregate is considered to be more anomalous if it contains many such infrequent unique message sequences. An information theoretic way to quantify the notion of infrequency is to compute the information content of the aggregate. Entropy is one such metric that succinctly quantifies information content. An aggregate with frequent unique message sequences will have less entropy due to less information content. On the other hand, an aggregate with more and more infrequent unique message sequences will have higher entropy due to higher  information content. The entropy of a message aggregate is {\em lower bounded} by 0.0 (when the aggregate contains exactly one unique message sequence) and is {\em upper bounded} by $log_2(n)$ (when the aggregate contains exactly one of each of the $n$ unique message sequences).

\begin{table}
    \caption{Definition of anomalies using engineered features entropy and Levenshtein distance. {\bf Ldist}: Levenshtein distance. \checkmark: Non-anomalous message aggregate. \xmark: Anomalous message aggregate.\label{table:ano_def}}
    \centering
    \begin{tabular}{|c||*{2}{c|}}
        \hline
        \backslashbox{\bf \em Ldist}{\bf \em Entropy}
        & \makebox[3em]{\bf Low} & \makebox[3em]{\bf High} \\
        \hline\hline
        {\bf Low} & \cellcolor{green}\checkmark & \cellcolor{green}\checkmark\\
        \hline
        {\bf High} & \cellcolor{green}\checkmark & \cellcolor{red}\textcolor{white}{\xmark} \\
        \hline
    \end{tabular}
\end{table}

\medskip

\noindent{\bf Levenshtein distance as an engineered feature}: Entropy fails to characterize the specific relationship that exists between individual unique message sequences of a message aggregate. Consequently, we calculate a {\em similarity metric}, in particular, Levenshtein distance (\cf,~\Cref{sec:ldistance}) to quantify the deviancy of the constituent message sequences in a message aggregate. If a message aggregate contains {\em similar} unique message sequences, the dissimilarity score will be small whereas if the message aggregate contains {\em deviant} unique message sequences, the dissimilarity score will be large. A message aggregate with higher Levenshtein distance will likely to be more anomalous as compared to another message aggregate with smaller Levenshtein distance. Levenshtein distance of a message aggregate is {\em lower bounded} by 0.0 (when the aggregate contains exactly one unique message sequence) and is {\em upper bounded} by the average of Hamming distance~\cite{cite:hamming} of pairwise unique message sequences (when the aggregate contains $n$ different unique message sequences).

\smallskip

Let us consider aggregates {\bf A1}: \{`aba', `bab'\} and {\bf A2}: \{`aba', `cdc'\} where a, b, c, d are messages. For each of the {\bf A1} and {\bf A2}, the entropy is $log_2(2) = 1$. Although {\bf A2} comprises dissimilar unique message sequences as compared to {\bf A1}, entropy alone fails to capture that dissimilarity. Hence we calculate the Levenshtein distance of each of the aggregates to quantify the dissimilarity of the constituent messages. For {\bf A1}, $\mathcal{L}$(`aba', `bab') = 2 (1 deletion and 1 insertion) and for {\bf A2}, $\mathcal{L}$(`aba', `cdc') = 3 (3 substitutions). Clearly, in spite of having same entropy, Levenshtein distance helped to identify {\bf A2} to be more anomalous than {\bf A1}.

In our diagnosis solution, we define a {\bf message aggregate as anomalous} (\ie, contains anomalous unique message sequences) that has {\bf both high entropy and high Levenshtein distance}.~\Cref{table:ano_def} summarizes our definition of anomalousness of a message aggregate.

\medskip

\noindent{\bf Usage of outlier detection algorithms}: We apply outlier detection algorithms to the engineered feature data space spanning over entropy and Levenshtein distance.  In the engineered feature space,
message aggregates that represents normal behavior will be very close to each other and will form a dense cluster. On the other hand, message aggregates that represents anomalous behavior will be sparsely distributed and distant from the normal message aggregates. Outlier algorithms output a ranked list of anomalous message aggregates ranked by outlier scores. We output message sequences contained in top-five anomalous message aggregates as candidate anomalies.

\begin{figure}
  \centering
  \includegraphics[scale=0.5]{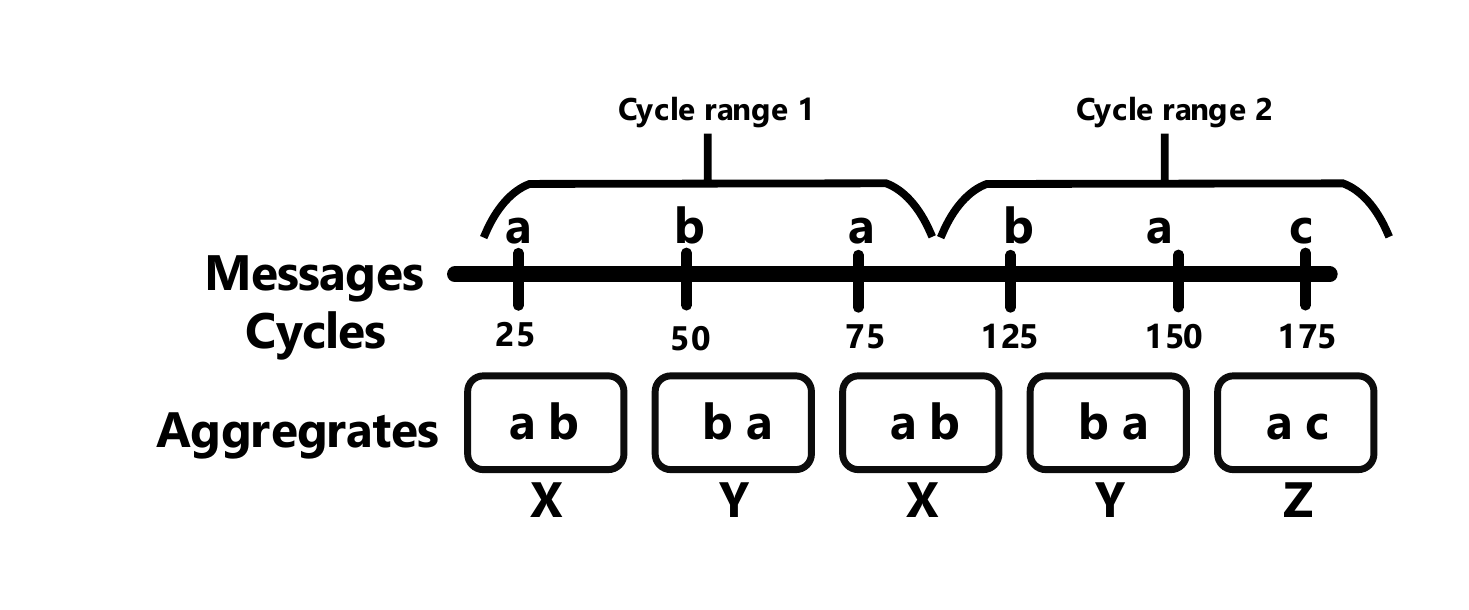}
  \caption{Example execution trace and a set of message sequences of length $k = 2$ and granularity $g = 100$ cycles.\label{fig:od_example}}
  \vspace{-1mm}
\end{figure}

\subsection{Example for generating engineered feature values from raw feature values}

We use an example trace of~\Cref{fig:od_example} to explain the steps for generating engineered feature values. This methodology is parameterized by i) the length $k$ of the message sequence for which anomaly needs to be detected and ii) the granularity $g$ in number of cycles at which message aggregates need to be created. For this example, we use $k = 2$ and $g = 100$.

\smallskip

\noindent {\bf Step 1 (Creation of message aggregates)}: We use a sliding window of length $k$ to create a set of $k$-length message sequences. The set of message sequences are partitioned into message aggregates based on granularity $g$. In the example, the set of two-length message sequences is S= \{$ab$, $ba$, $ab$, $ba$, $ac$\}. We partition S at a granularity of 100 cycles which creates two message aggregates $s_1 = \{X, Y, X\}$ and $s_2 = \{X, Y, Z\}$ where $X = ab, Y = ba, Z = ac$.

\medskip

\noindent {\bf Step 2 (Identifying unique message sequences and their occurrences per message aggregate)}: We identify unique message sequences per message aggregate and calculate their number of occurrences. In this example, $s_1$ has two unique message sequences $X$ and $Y$, and $s_2$ has three unique message sequences $X$, $Y$, and $Z$. In $s_1$, $X$ happened two times, and $Y$ happened one time. In $s_2$, each of the $X$, $Y$, and $Z$ has happened one time.

\medskip

\noindent {\bf Step 3 (Calculation of entropy and Levenshtein distance per message aggregate)}: We calculate entropy and Levenshtein distance for each of the message aggregates using the information of unique message sequences from Step 2.

In the example, for aggregate $s_1$, $p(X) = 2/3$ and $p(Y) = 1/3$. Hence $\mathcal{H}(s_1) = -p(X) log_2(X) - p(Y) log_2(Y) = -2/3 * log_2(2/3) - 1/3 * log_2 (1/3) = 0.9182$ and $\mathcal{L}(X, Y) = 2,~\mathcal{L}(X, X) = 0$, and $\mathcal{L}(Y, X) = 2$. The average Levenshtein distance of aggregate $s_1$ is $(2 + 0 + 2)/3 = 1.33$.

Similarly, for aggregate $s_2$,  $p(X) = 1/3$, $p(Y) = 1/3$, and $p(Z) = 1/3$. Hence $\mathcal{H}(s_2) = -~p(X) log_2(X) - p(Y) log_2(Y) - p(Z) log_2(Z)= -~1/3 * log_2(1/3) - 1/3 * log_2 (1/3) - 1/3 * log_2 (1/3) = 1.58$ and $\mathcal{L}(X, Y) = 2,~\mathcal{L}(X, Z) = 2$, and $\mathcal{L}(Y, Z) = 2$. The average Levenshtein distance of aggregate $s_2$ is $(2 + 2 + 2)/3 = 2.0$.

The aggregates $s_1$ and $s_2$ are represented by tuples (0.9182, 1.33) and (1.58, 2.0) respectively in engineered feature space. We input these tuples to outlier detection algorithms to detect anomalous message aggregates.


\subsection{Limitations of the engineered features}\label{sec:scope_eng_feature}


\deb{Our proposed engineered features can diagnose a wide range of post-silicon use-case failures. 
However we do not claim that our features are sufficient to diagnose any arbitrary post-silicon use-case failures. For instance, if a buggy post-silicon execution trace consists of a very few unique messages repetitively (\eg,~`abcabcabc\ldots' where a, b, and c are unique messages), then the message aggregates will have frequently occurring similar unique message sequences. This will result in {\em low entropy} (due to frequent occurrence of each of the unique message sequences in the aggregate) and {\em small average Levenshtein distance} per message aggregate (due to similar unique message sequences in the aggregate) causing our method to fail to diagnose the bug. Certain class of bugs may escape our diagnosis method if the engineered features fail to demarcate correct and buggy behaviors in the engineered feature space. However this does not limit the practical applicability of our diagnosis solution. Our solution expedites predominantly manual post-silicon debugging by several orders of magnitude (\cf,~\Cref{sec:comprehensive}). To make our solution comprehensive, additional engineered features are needed.}

\begin{figure}
	\centering
	\includegraphics[scale=0.23]{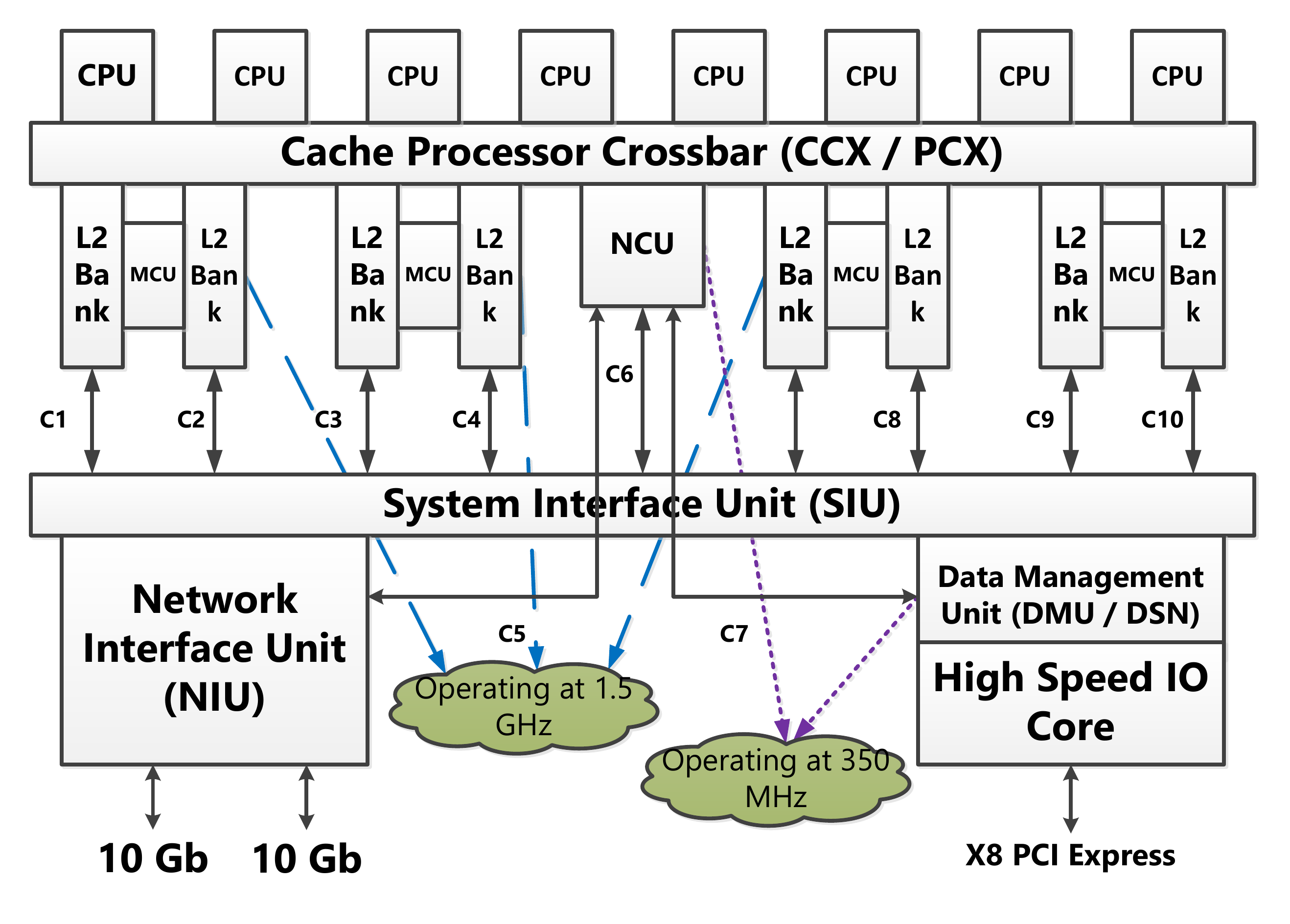}
	\caption{Block diagram of OpenSPARC T2 processor. {NCU}: Non-cacheable unit, {MCU}: Memory Controller Unit~\cite{cite:SPARCT2Vol1, cite:SPARCT2Vol2}\label{fig:opensparcblock}}
	\vspace{-4mm}
\end{figure}

\section{Experimental Setup}\label{sec:ExpSet}

\noindent {\bf \em Design testbed}: We primarily use the publicly available OpenSPARC T2 SoC~\cite{cite:SPARCT2Vol1, cite:SPARCT2Vol2} to demonstrate our result. Figure~\ref{fig:opensparcblock} shows an IP level block diagram of T2. Three different usage scenarios considered in our debugging case studies are shown in Table~\ref{table:usagescen} along with participating flows (column 2-6) and participating IPs (column 7). We also use the USB design~\cite{cite:USB} to compare with other methods that cannot scale to the T2.

\begin{figure}
\centering
	\includegraphics[scale=0.5]{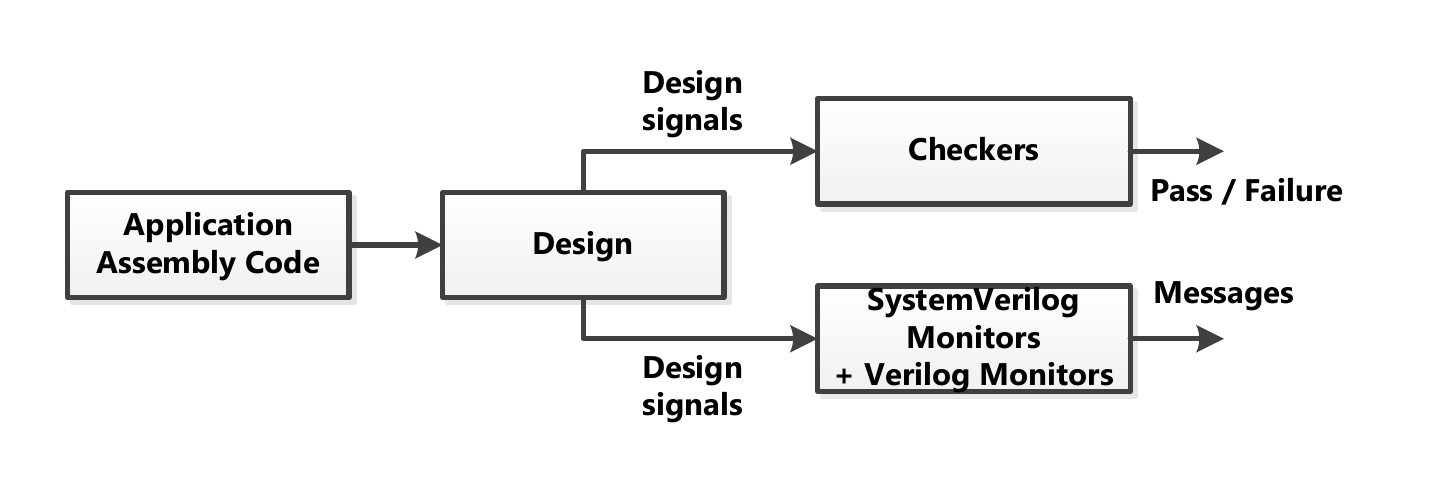}
	\caption{Experimental setup to convert design signals to flow messages\label{fig:rtl_sig_to_flow_msg}}
	\vspace{-1mm}
\end{figure}

\begin{table}
    \scriptsize
    \caption{\small Usage scenarios and participating flows in T2. {\bf UID}: Usage scenario ID. {\bf PI}: participating IPs. {\bf PRC}: Number of potential root causes. {\bf PIOR:} PIO read, {\bf PIOW:} PIO write, {\bf NCUU:} NCU upstream, {\bf NCUD:} NCU downstream and {\bf Mon:} Mondo interrupt flow. \checkmark indicates Scenario i executes a flow $j$ and \xmark~indicates Scenario i does not execute a flow $j$. Flows are annotated with (No of flow states, No of messages).\label{table:usagescen}}
    \centering
    \begin{tabular}{|>{\centering}m{0.4cm}|>{\centering}m{0.55cm}|
    >{\centering}m{0.55cm}|>{\centering}m{0.6cm}|>{\centering}m{0.6cm}|
    >{\centering}m{0.55cm}|>{\centering}m{1.5cm}|c|}
	\hline
	\multirow{2}{*}{\bf UID} & \multicolumn{5}{|c|}{\bf Participating flows} & \multirow{2}{*}{\bf PI} & \multirow{2}{*}{\bf PRC}\\
	\cline{2-6}
	{} & {\footnotesize PIOR \scriptsize{(6, 5)}} & {\footnotesize PIOW \scriptsize{(3, 2)}} & {\footnotesize NCUU \scriptsize{(4, 3)}} & {\footnotesize NCUD \scriptsize{(3, 2)}} & {\footnotesize Mon \scriptsize{(6, 5)}} & {} & {}\\
	\hline\hline
	\sceneone  & \checkmark & \checkmark & \xmark & \xmark & \checkmark & \scriptsize{NCU, DMU, SIU} & 9 \\
	\hline
	\scenetwo  & \xmark & \xmark & \checkmark & \checkmark & \checkmark & \scriptsize{NCU, MCU, CCX} & 8 \\
	\hline
	\sceneth   & \checkmark & \checkmark & \checkmark & \checkmark & \xmark & \scriptsize{NCU, MCU, DMU, SIU} & 9\\
	\hline
\end{tabular}
\end{table}

\begin{table}
	\centering
	\scriptsize
	\caption{Representative bugs injected in IP blocks of OpenSPARC T2. {\bf Bug depth}
indicates the hierarchical depth of an IP block from the top. Bug type is the functional implication of a bug. \label{table:bugdetails}}
    \begin{tabular}{|>{\centering}m{0.3cm}|>{\centering}m{0.5cm}|>{\centering}m{0.8cm}|
    >{\centering}m{4.5cm}|p{0.5cm}|}
		\hline
		{\bf Bug} & {\bf Bug} & {\bf Bug} & {\bf Bug} & {\bf Buggy}\\
		{\bf ID} & {\bf depth} & {\bf category} & {\bf type} & {\bf IP}\\
		\hline\hline
		1 & 4 & Control & wrong command generation by data misinterpretation & DMU \\ 
		\hline
		2 & 4 & Data    & Data corruption by wrong address generation & DMU\\ 
		\hline
		3 & 3 & Control & Wrong construction of Unit Control Block resulting in malformed request & DMU\\ 
		\hline
		4  & 4 & Control & Generating wrong request due to incorrect decoding of request packet from CPU buffer & NCU\\
		\hline
	\end{tabular}
\end{table}

\smallskip

\noindent {\bf \em Testbenches}: We used 37 different tests from {\tt fc1\_all\_T2} regression environment. Each test exercises two or more IPs and associated flows. We monitored message communication across participating IPs and recorded the messages into an output trace file using the System-Verilog monitor of~\Cref{fig:rtl_sig_to_flow_msg}. We also record the status (passing/failing) of each of the tests.

\smallskip

\noindent {\bf \em Bug injection}: We created 5 different buggy versions of T2, that we analyze as five different case studies. Each case study comprises 5 different IPs. We injected a total of 14 different bugs across the 5 IPs in each case. The injected bugs follow two sources~\textendash i) sanitized examples of communication bugs received from our industrial partners and ii) the ``bug model" developed at the Stanford University in the QED~\cite{lin2014effective} project capturing commonly occurring bugs in an SoC design. A few representative injected bugs are detailed in Table~\ref{table:bugdetails}. Table~\ref{table:bugdetails} shows that the set of injected bugs are complex, subtle and realistic. It took up to {\bf 457 observed messages} and up to {\bf 21290999 clock cycles} for each bug symptom to manifest. These demonstrate complexity and subtlety of the injected bugs. Following~\cite{cite:SPARCT2Vol1, cite:SPARCT2Vol2} and Table~\ref{table:bugdetails}, we have identified several potential architectural causes that can cause an execution of a usage scenario to fail. Column 8 of Table~\ref{table:usagescen} shows number of potential root causes per usage scenario.

\smallskip

\noindent {\bf \em Anomaly detection techniques}: We used six different outlier detection  algorithms, namely IForest, PCA, LOF, LkNN (kNN with longest distance method), MukNN (kNN with mean distance method), and OCSVM from PyOD~\cite{cite:zhao2019pyod}. We applied each of the above outlier detection algorithms on the failure trace data generated from each of the five different case studies to diagnose anomalous message sequences that are symptomatic of each of the injected bugs per case study. 

\begin{table}
	\centering
	\scriptsize
	\caption{Trace buffer utilization flow specification coverage and path localization of traced messages
     for 3 different usage scenarios. {\bf FSP Cov}: Flow specification coverage (Definition~\ref{def:flow_spec_cov}), WP: With packing,  WoP: Without Packing. 32 bits wide trace buffer assumed. \label{table:TraceBufUtil}}
    \begin{tabular}{|>{\centering}m{0.6cm}|>{\centering}m{1.1cm}|>{\centering}m{0.6cm}|>{\centering}m{0.6cm}|>{\centering}m{0.6cm}|>{\centering}m{0.6cm}|>{\centering}m{0.6cm}|p{0.6cm}|}
		\hline
		{\bf Case} & {\bf Usage} & \multicolumn{2}{|c|}{\bf Trace Buffer} & \multicolumn{2}{|c|}{\bf FSP Cov} & \multicolumn{2}{|c|}{\bf  Path}\\
        {\bf  study} & {\bf Scenario} & \multicolumn{2}{|c|}{\bf  Utilization} &\multicolumn{2}{|c|}{} & \multicolumn{2}{|c|}{\bf  Localization} \\
		\cline{3-8}
		& & WP & WoP & WP & WoP & WP & WoP\\
		\hline
		\hline
		1 & \multirow{3}{*}{\sceneone} & \multirow{3}{*}{96.88\%} & \multirow{3}{*}{84.37\%} &
		\multirow{3}{*}{99.86\%} & \multirow{3}{*}{97.22\%} &  0.13\% &
        3.23\%\\
		\cline{7-8}
        \cline{1-1}
		2 & &    &   &   &  &  0.31\% & 6.11\% \\
		\hline
		3 & \multirow{2}{*}{\scenetwo} & \multirow{2}{*}{100\%} & \multirow{2}{*}{71.87\%} &
        \multirow{2}{*}{99.69\%} & \multirow{2}{*}{93.75\%}  &  0.26\% & 5.13\% \\
		\cline{7-8}
        \cline{1-1}
		4 & &    &   &   &    &  0.10\% & 2.47\% \\
		\hline
		5 & \sceneth & 100\% & 93.75\% & 83.33\% & 77.78\% &  0.11\% & 2.65\%\\
		\hline
	\end{tabular}
\end{table}

\section{Experimental results on message selection}\label{sec:exp_result_selection}

In this section, we provide insights into the effectiveness of our message selection technique using
five different (buggy) case studies across 3 usage scenarios of the T2.

\begin{figure}
	\centering
	\includegraphics[scale=0.3]{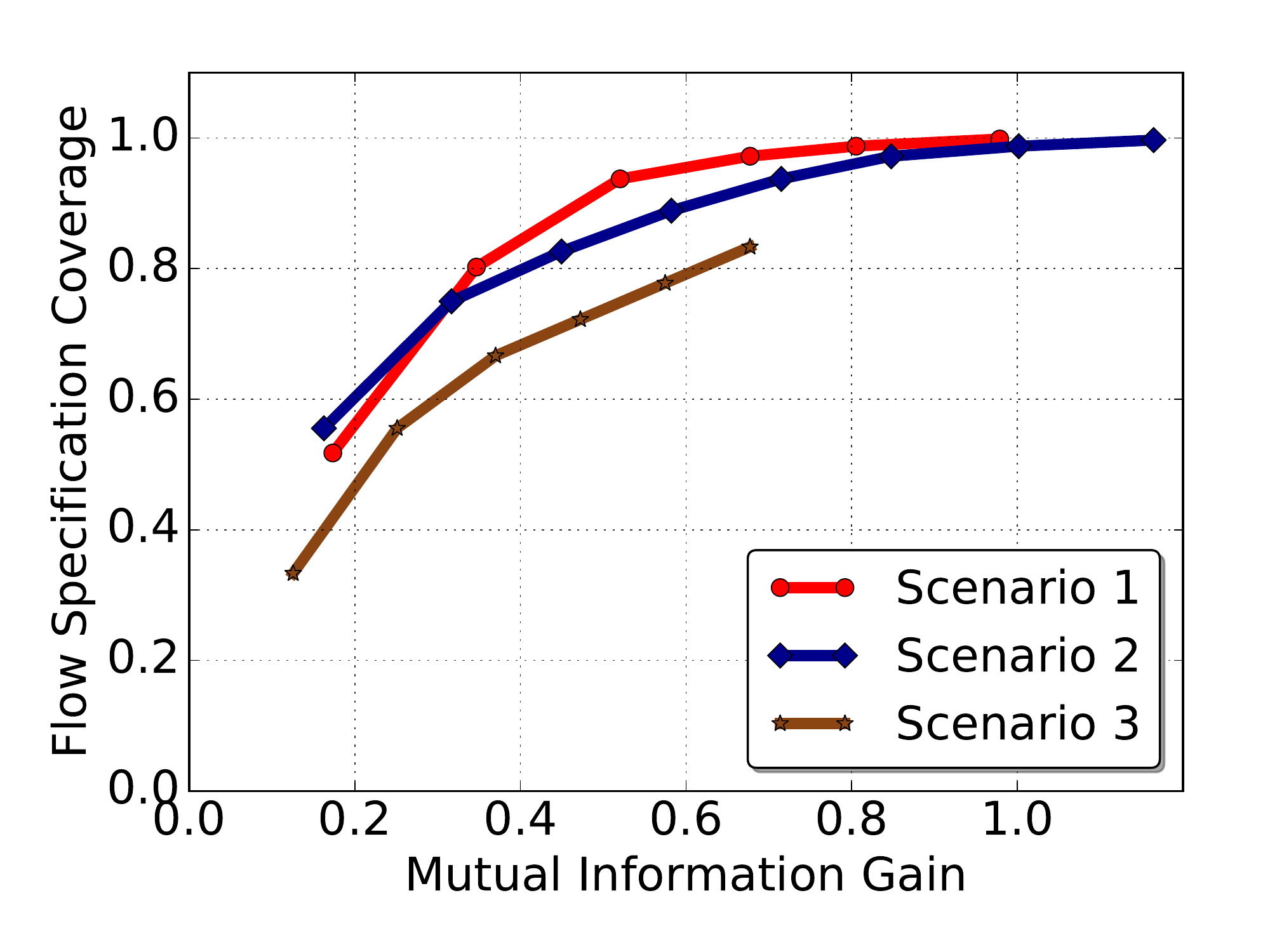}
	\caption{Correlation analysis between {\em mutual information gain} and {\em flow
     specification coverage} for different message combinations for three different usage scenarios.\label{fig:corrmutualinfo}}
     \vspace{-2mm}
\end{figure}

\subsection{Flow specification coverage and trace buffer utilization}\label{sec:funcspeccov}

Table~\ref{table:TraceBufUtil} demonstrates the value of the traced messages with respect to flow specification coverage (Definition~\ref{def:flow_spec_cov}) and trace buffer utilization. These are the two objectives for which our message selection is optimized. Messages selected {\bf without packing} achieve {\bf up to 93.75\% of trace buffer utilization} with {\bf up to 97.22\% flow specification coverage}. {\bf With packing}, message selection achieves {\bf up to 100\% of trace buffer utilization} and {\bf up to 99.86\% flow specification coverage}. This shows that we can cover most of the desired functionality while utilizing the trace buffer maximally.

\begin{table}
	\centering
	\scriptsize
	\caption{Comparison of signals selected by our method with SigSeT~\cite{cite:basu2011efficient} and
		PRNet~\cite{cite:conf/iccad/MaPJRV15} for the USB design. {\bf P}: Partial bit
		\label{table:USBSelcetedSignals}}
	\begin{tabular}{|c|c|c|c|c|}
		\hline
		{\bf Signal}	& {\bf USB}	&  {\bf Sig} & {\bf PR}  & {\bf Info}\\
		{\bf Name}      & {\bf Module}    & {\bf SeT} & {\bf Net} & {\bf Gain}\\
		\hline\hline
		rx\_data & UTMI & \xmark & \checkmark & \checkmark \\
		\cline{1-1}\cline{3-5}
		rx\_valid & line speed & \xmark & \checkmark & \checkmark\\
		\cline{1-1}\cline{3-5}
		\hline
		rx\_data \_valid & Packet & \xmark & \xmark & \checkmark\\
		\cline{1-1}\cline{3-5}
		token \_valid & decoder & \xmark & \xmark & \checkmark\\
		\cline{1-1}\cline{3-5}
		rx\_data \_done & & \xmark & \xmark & \checkmark\\
		\hline		
		tx\_data & Packet & \xmark & \xmark & \checkmark\\
		\cline{1-1}\cline{3-5}
		tx\_valid & assembler & \xmark & \checkmark & \checkmark\\
		\cline{1-1}\cline{3-5}
		\hline
		send\_token & Protocol & \xmark & \xmark & \checkmark\\
		\cline{1-1}\cline{3-5}
		token\_pid \_sel & engine & {\bf P} & {\bf P} & \checkmark\\
		\cline{1-1}\cline{3-5}
		data\_pid \_sel & & {\bf P} & \xmark & \checkmark\\
		\hline
	\end{tabular}
\end{table}

\subsection{Path localization during debug of traced messages}\label{sec:pathlocal}
In this experiment, we use buggy executions and traced messages to show the extent of path localization per bug. Localization is calculated as the fraction of total paths of the interleaved flow. In Table~\ref{table:TraceBufUtil}, columns 7 and 8 show the extent of path localization. We needed to explore {\bf no more than 6.11\% of interleaved flow paths} using our selected messages. With packing, we needed to explore {\bf no more than 0.31\% of the total interleaved flow paths} during debugging. Even with packing, subtle bugs like NCU bug of buggy design 3~and buggy design 2~needed more paths to explore.

\begin{table}
  \centering
  \scriptsize
  \caption{Selection of important messages by our method\label{table:InjBugMsgSel}}
  \begin{tabular}{|c|c|c|c|c|c|}
    \hline
    {\bf Message} & {\bf Affecting} & {\bf Bug} & {\bf Message} & \multicolumn{2}{|c|}{\bf Selected} \\
     \cline{5-6}
     & {\bf Bug IDs} & {\bf coverage} & {\bf importance} & {\bf Y / N } & {\bf Usage}\\
     &  &  &  &  & {\bf scenario}\\
    \hline
    {\bf m1} & 8, 33, 36 & 0.21 & 4.76 & Y & 1, 2 \\
    \hline
    {\bf m2} & 8, 33, 34, 36 & 0.28 & 3.57 & Y & 1, 2 \\
    \hline
    {\bf m3} & 33, 36 & 0.14 & 7.14 & Y & 1, 2 \\
    \hline
    {\bf m4} & 8, 29, 33 & 0.21 & 4.76 & Y & 1, 3 \\
    \hline
    {\bf m5} & 18, 33 & 0.14 & 7.14 & Y & 1, 2 \\
    \hline
    {\bf m6} & - & - & & N & - \\
    \hline
    {\bf m7} & - & - & & Y & 1, 3 \\
    \hline
    {\bf m8} & 33 & 0.07 & 14.28 & Y & 2 \\
    \hline
    {\bf m9} & 1, 33 & 0.14 & 7.14 & N & - \\
    \hline
    {\bf m10} & 24 & 0.07 & 14.28 & Y & 2 \\
    \hline
    {\bf m11} & 1, 24 & 0.14 & 7.14 & Y & 2 \\
    \hline
    {\bf m12} & 24 & 0.07 & 14.28 & Y & 2 \\
    \hline
    {\bf m13} & 8 & 0.07 & 14.28 & Y & 2 \\
    \hline
    {\bf m14} & 1, 17, 33 & 0.21 & 4.76 & Y & 2 \\
    \hline
    {\bf m15} & 1, 17, 18, 33 & 0.28 & 3.57 & N & - \\
    \hline
    {\bf m16} & 1, 17, 18, 33 & 0.28 & 3.57 & Y & 2, 3 \\
    \hline
  \end{tabular}
\end{table}

\begin{figure}
    \centering
    \includegraphics[scale=0.28]{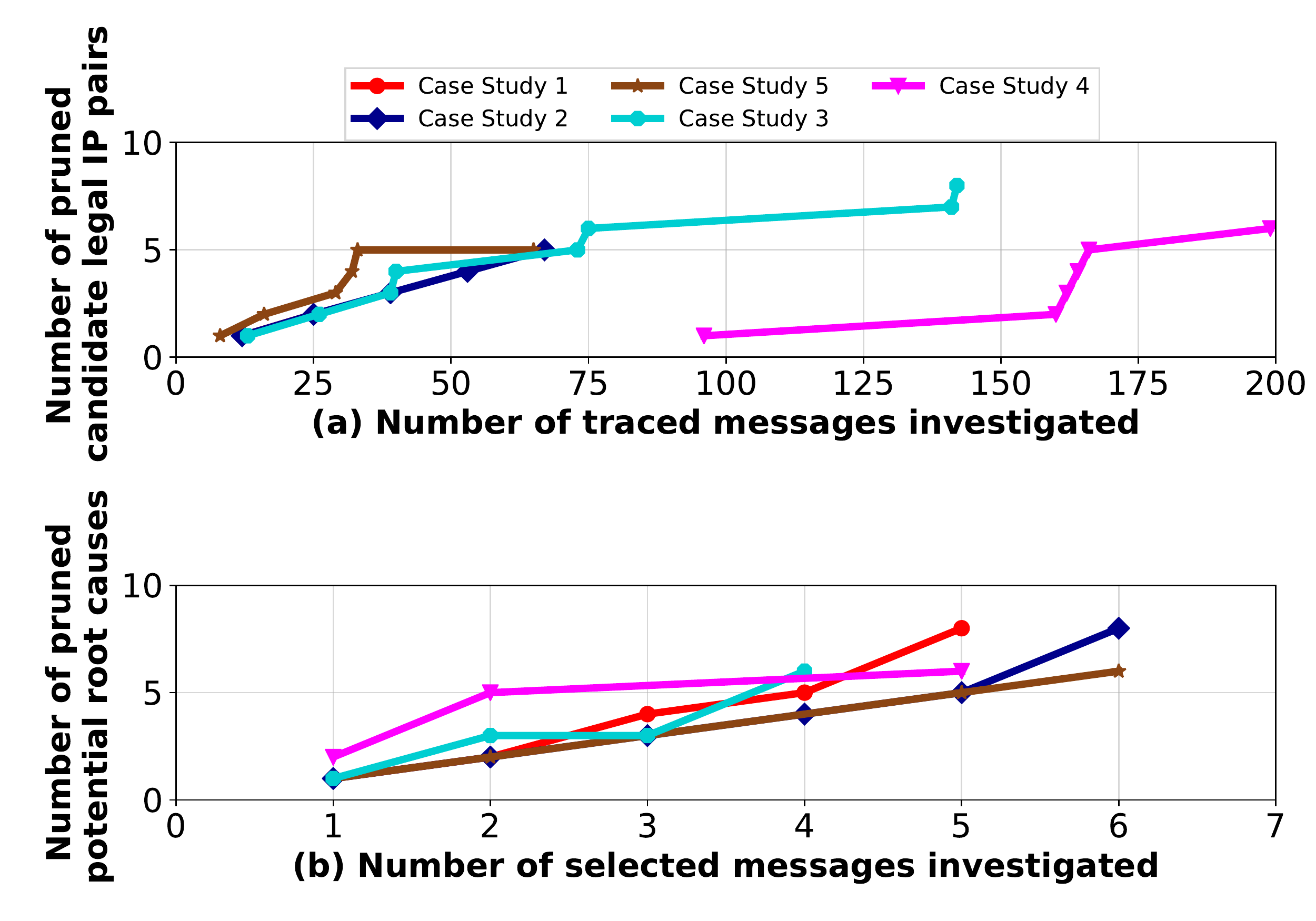}
    \caption{Root causing buggy IP\label{fig:cumulative_cause_eli}}
\end{figure}

\begin{figure}
	\centering
	\begin{subfigure}[b]{0.15\textwidth}
		\includegraphics[scale=0.11]{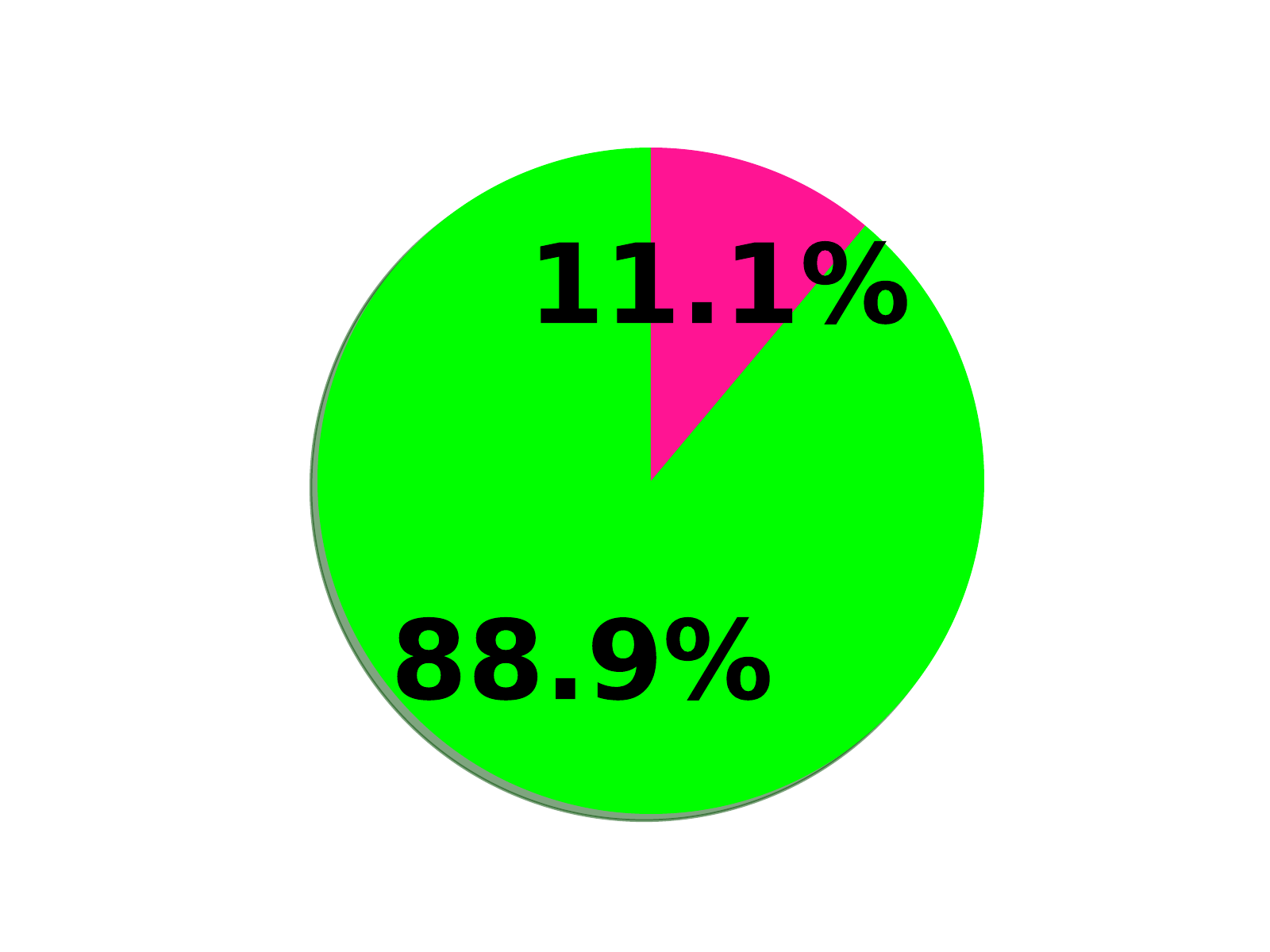}
		\caption{\footnotesize Case Study 1}
		\label{fig:des1scene1}
	\end{subfigure}
	\hfill
	\begin{subfigure}[b]{0.15\textwidth}
		\includegraphics[scale=0.11]{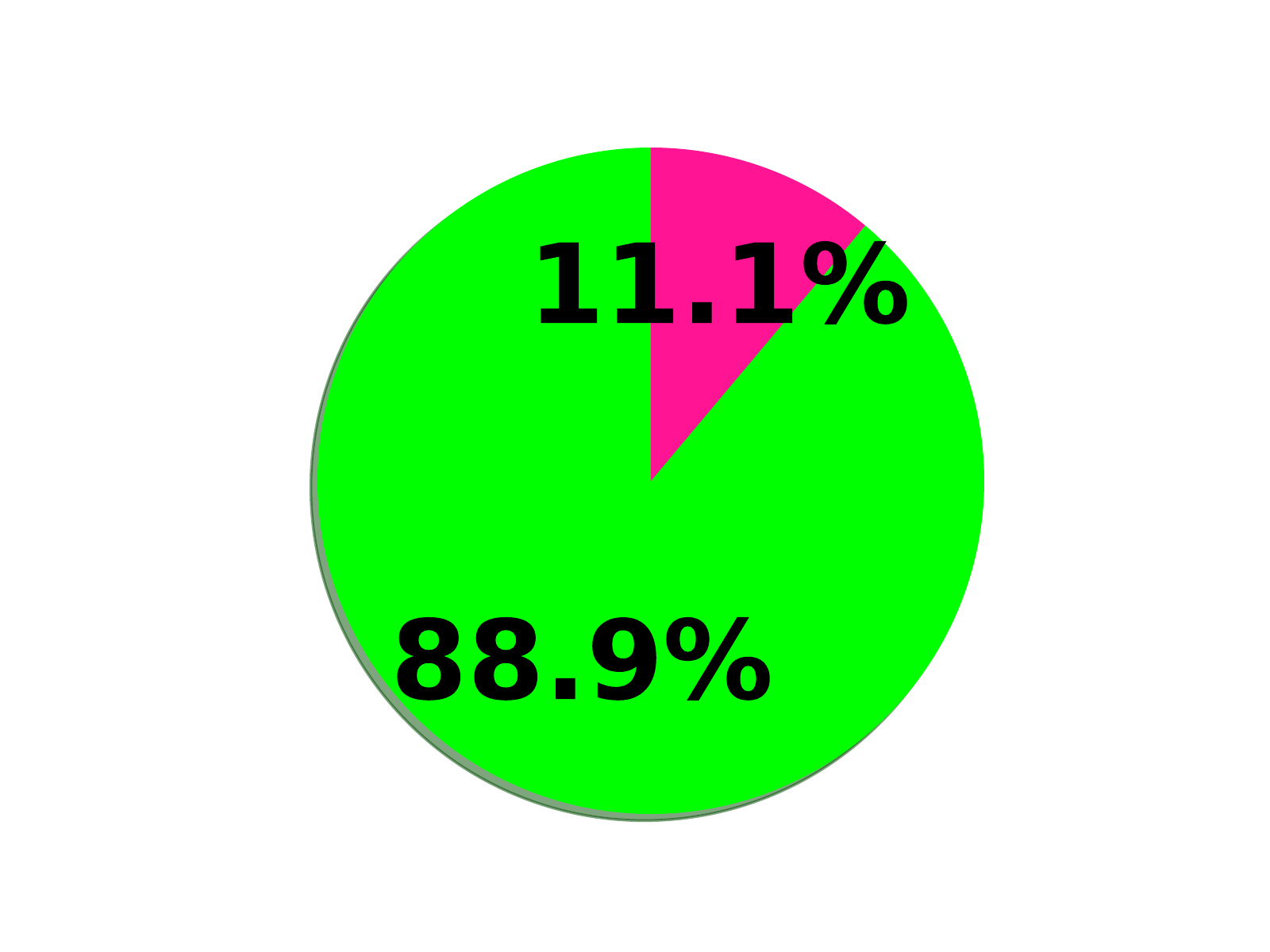}
		\caption{\footnotesize Case Study 2}
		\label{fig:des3scene1}
	\end{subfigure}
	\hfill
	\begin{subfigure}[b]{0.15\textwidth}
		\includegraphics[scale=0.11]{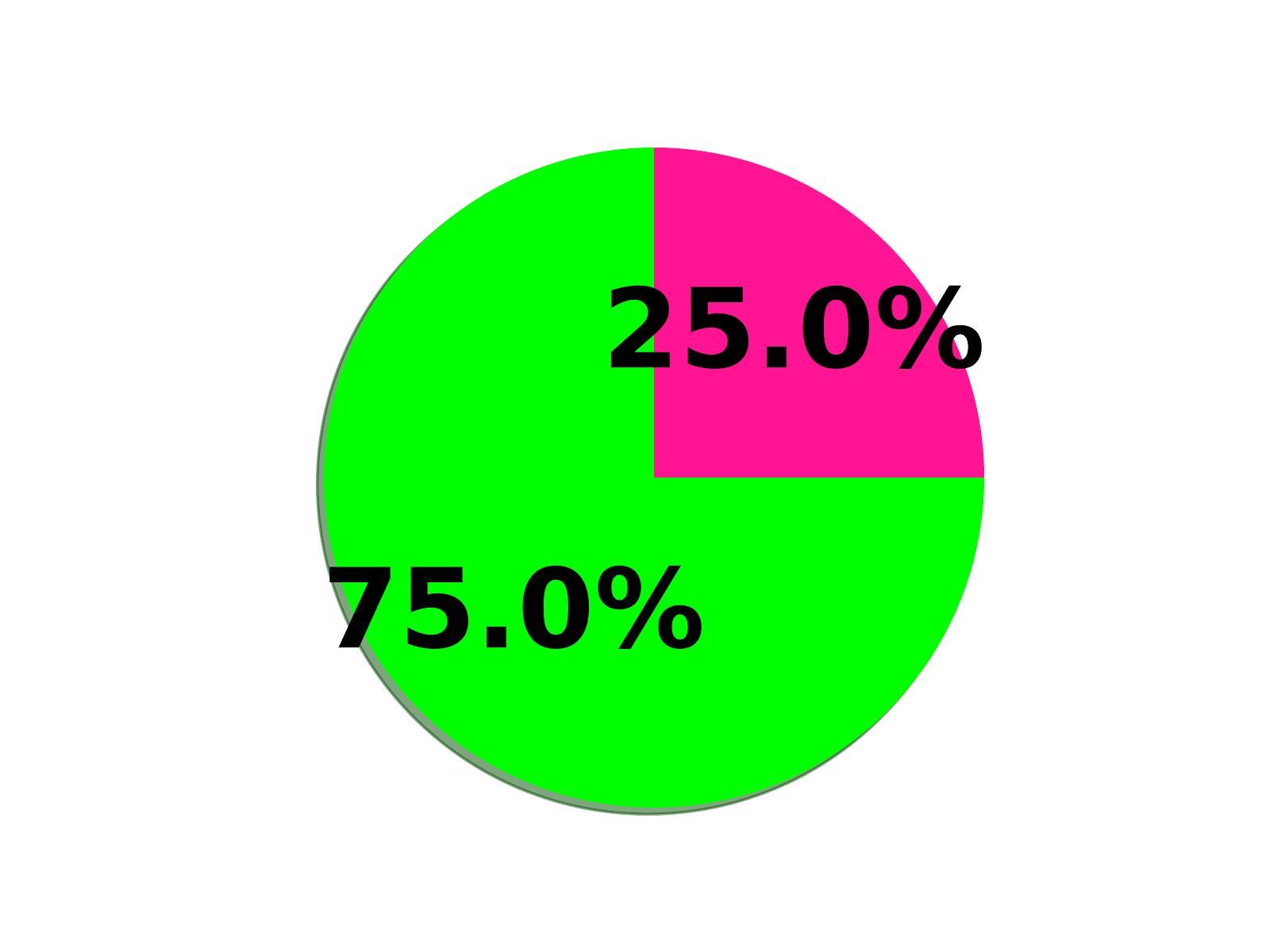}
		\caption{\footnotesize Case Study 3}
		\label{fig:des1scene2_2}
	\end{subfigure}
	\hfill
	\begin{subfigure}[b]{0.15\textwidth}
		\includegraphics[scale=0.11]{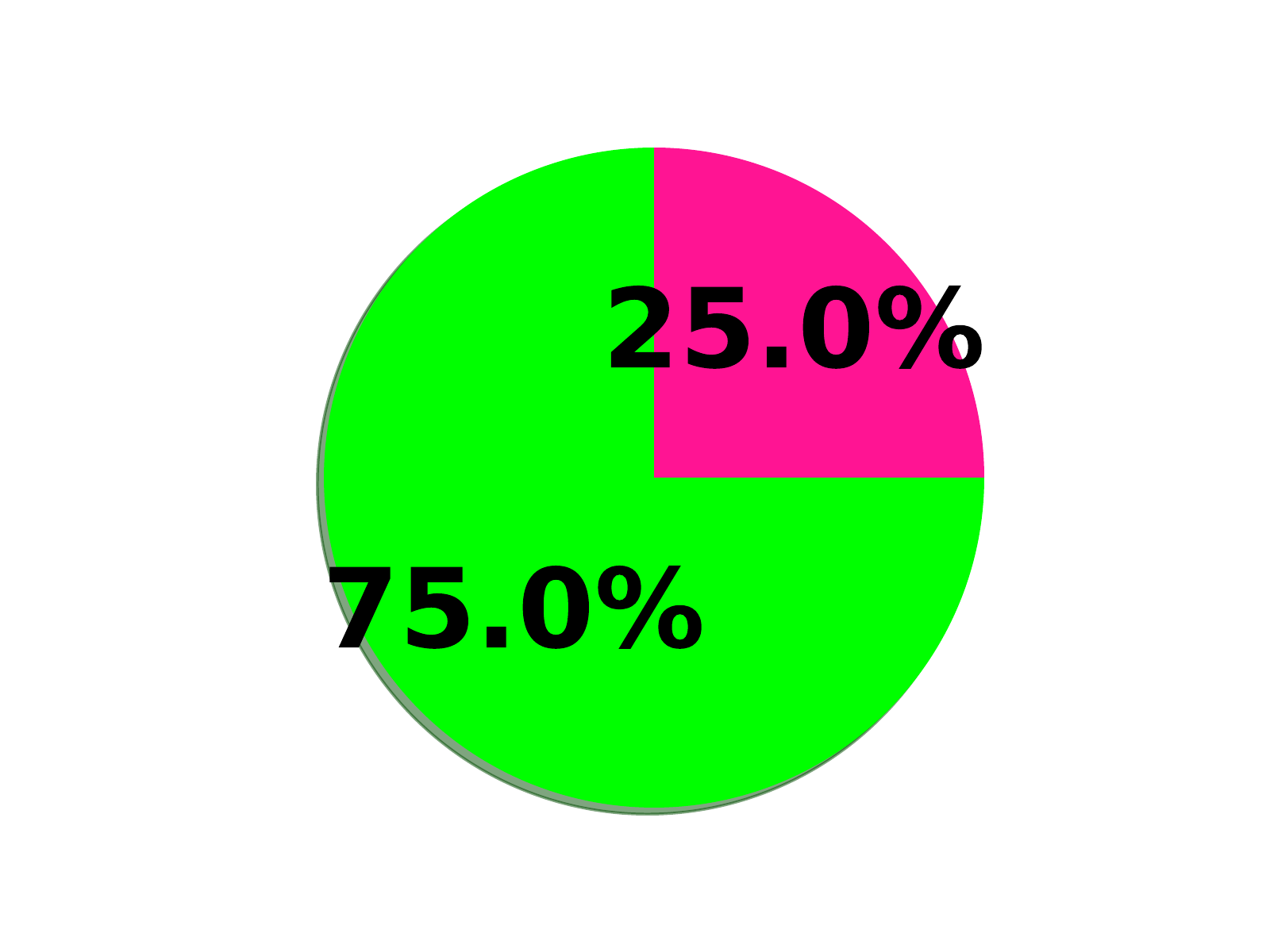}
		\caption{\footnotesize Case Study 4}
		\label{fig:des2scene2}
	\end{subfigure}
	\begin{subfigure}[b]{0.15\textwidth}
		\includegraphics[scale=0.11]{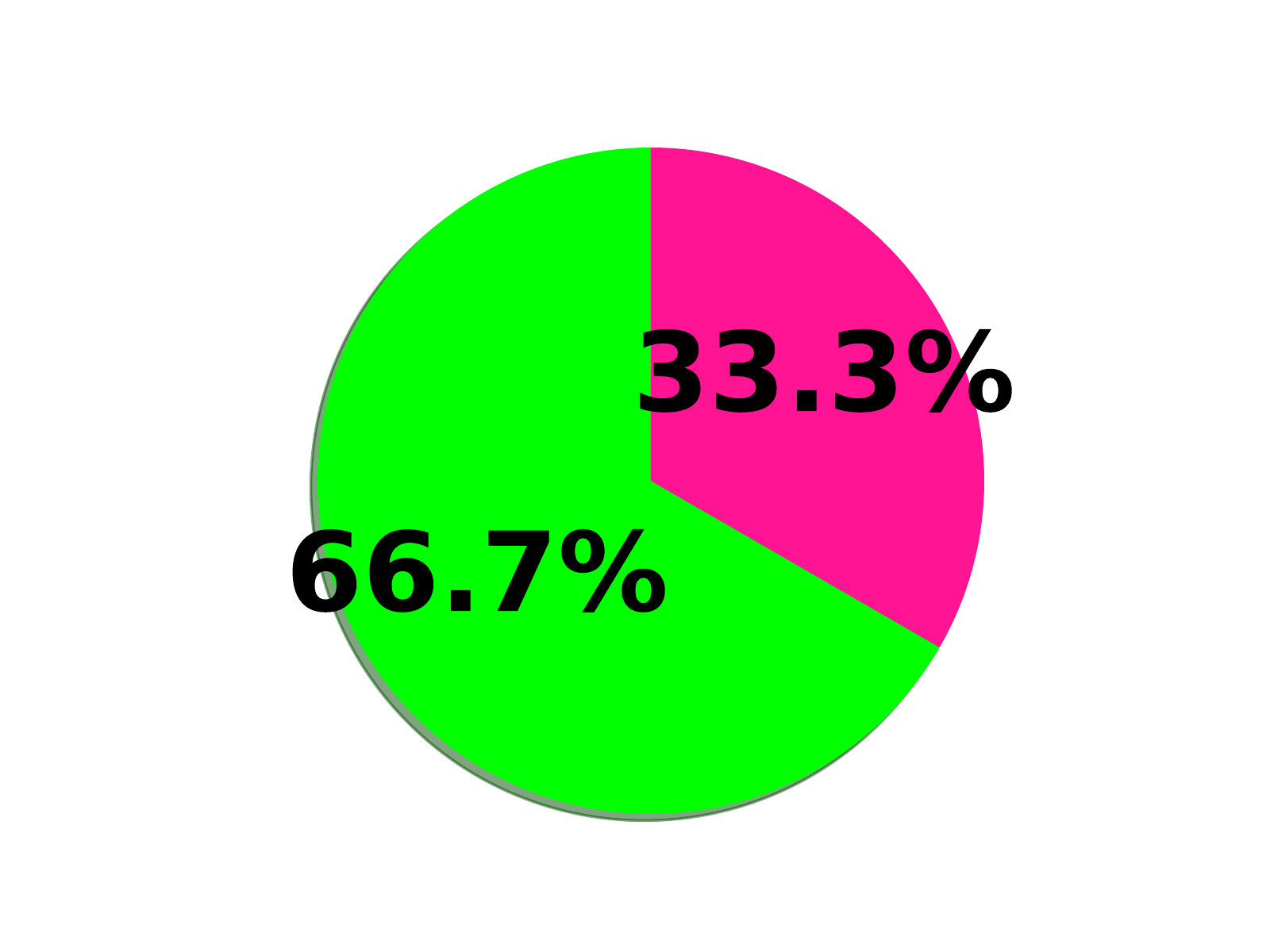}
		\caption{\footnotesize Case Study 5}
		\label{fig:des5scene3}
	\end{subfigure}
	\caption{Selected messages-cause pruning distribution for diagnosis.
		\fcolorbox{black}{magenta}{\rule{0pt}{2pt}\rule{2pt}{0pt}}~Plausible Cause,
		\fcolorbox{black}{green}{\rule{0pt}{2pt}\rule{2pt}{0pt}}~Pruned Cause
		\label{fig:CauseDistrib}}
\end{figure}

\subsection{Validity of information gain as message selection metric}\label{sec:infogainspeccov}

We select messages per usage scenario. In Figure~\ref{fig:corrmutualinfo} we analyze the correlation between flow specification coverage and the mutual information gain of the selected messages. Flow specification coverage (Definition~\ref{def:flow_spec_cov}) {\bf increases monotonically with the mutual information gain} over the interleaved flow of the corresponding usage scenario. This establishes that {\bf increase in mutual information gain corresponds to higher coverage of flow specification}, indicating that mutual information gain is a good metric for message selection.

\begin{table*}
	\centering
	\scriptsize
	\caption{Diagnosed root causes and debugging statistics for our case studies on OpenSPARC T2. \label{table:DebuggingSteps}}
	\begin{tabular}{|>{\centering}m{1.2cm}|>{\centering}m{0.5cm}|>{\centering}m{1.2cm}|>{\centering}m{1.2cm}|
        >{\centering}m{1.5cm}|p{9.5cm}|}
		\hline
		{\bf Case Study} & {\bf No of. Flows} & {\bf Legal IP Pairs} & {\bf
			Legal IP pairs investigated} & {\bf Messages investigated} & {\bf Root caused architecture level function} \\
		\hline
		1 & 3 & \multirow{2}{*}{12} & 5 & 25 & An interrupt was never generated
		by DMU due to wrong interrupt generation logic\\
		\cline{1-2}\cline{4-6}
		2 & 3 & & 6 & 67 & Wrong interrupt decoding logic in NCU / Corrupted interrupt handling table in
		NCU\\
		\hline
		3 & 3 & \multirow{2}{*}{10} & 8 & 142 & Malformed CPU request from
		Cache Crossbar to NCU / Erroneous CPU request decoding logic of NCU\\
		\cline{1-2}\cline{4-6}
		4 & 3 & & 6 & 199 & Erroneous interrupt deque logic after interrupt is serviced\\
		\hline
		5 & 4 & 12 & 5 & 65 &Erroneous decoding logic of CPU requests in memory controller\\
		\hline
	\end{tabular}
\end{table*}

\begin{figure*}
	\centering
	\begin{subfigure}[b]{0.32\textwidth}
        \includegraphics[scale=0.40]{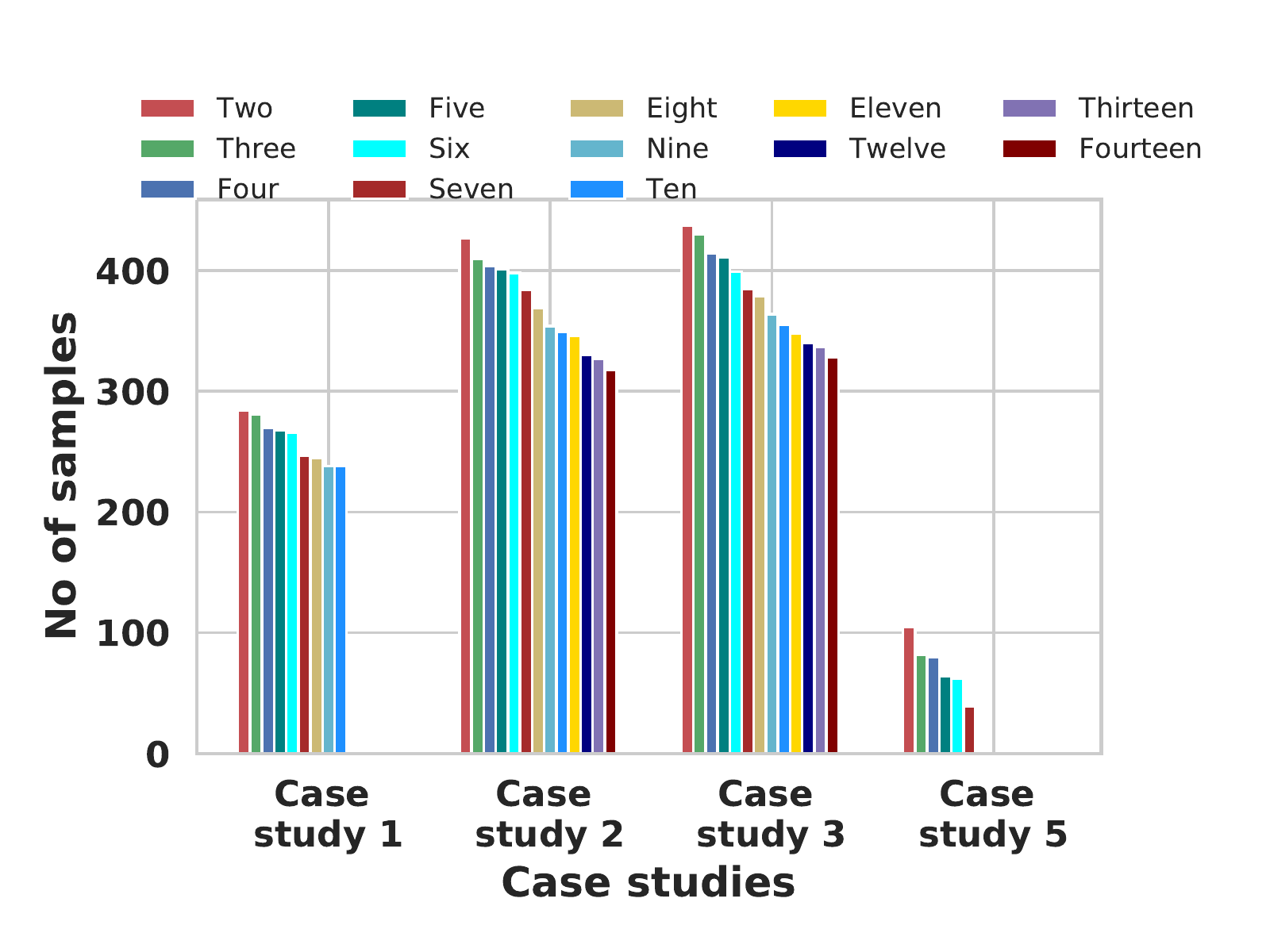}
		\caption{Total number of samples\label{fig:sample_num}}
	\end{subfigure}
	\hfill
	\begin{subfigure}[b]{0.32\textwidth}
        \includegraphics[scale=0.40]{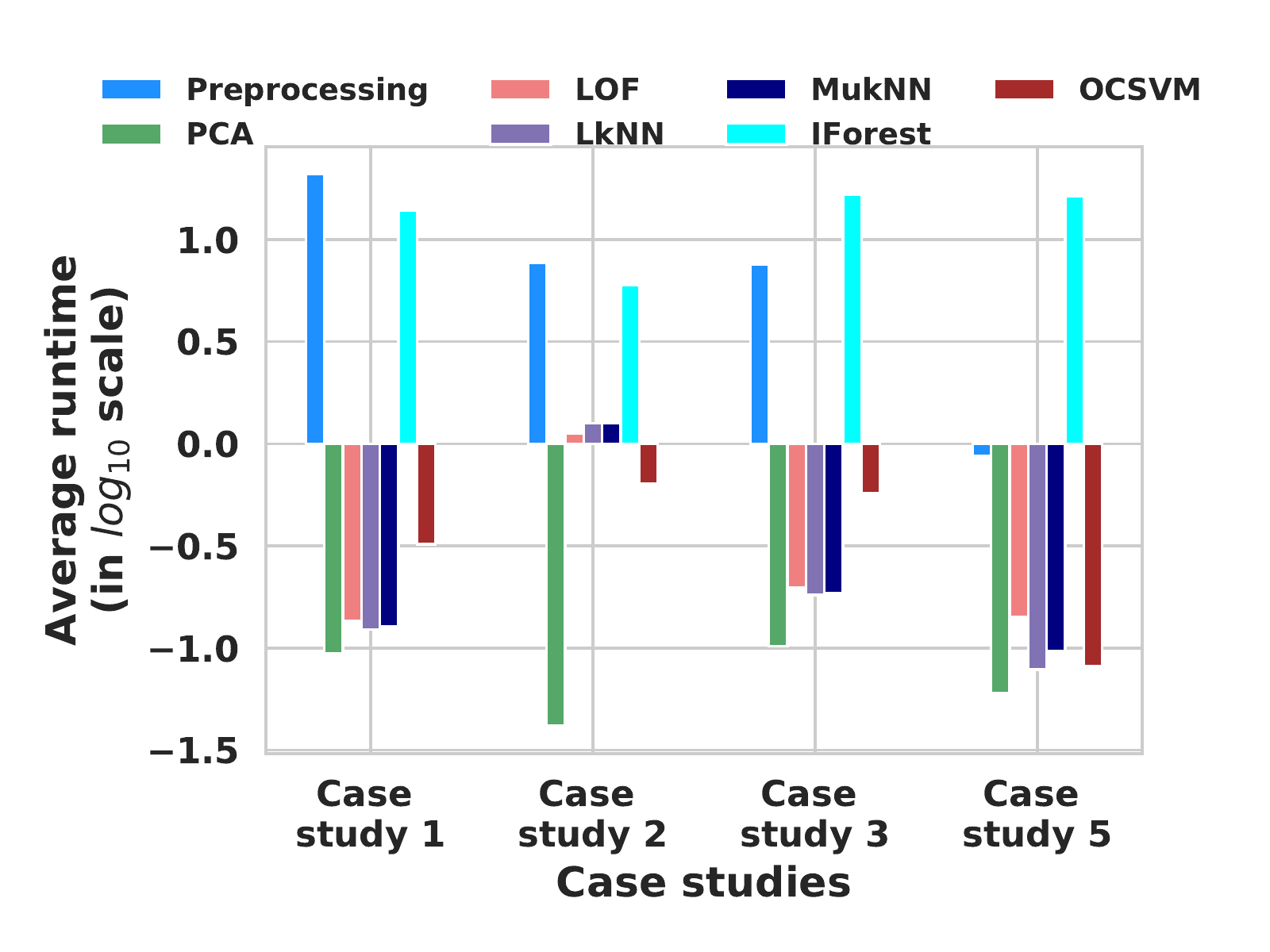}
		\caption{Run time comparison\label{fig:ad_runtime}}
	\end{subfigure}
	\hfill
	\begin{subfigure}[b]{0.32\textwidth}
        \includegraphics[scale=0.40]{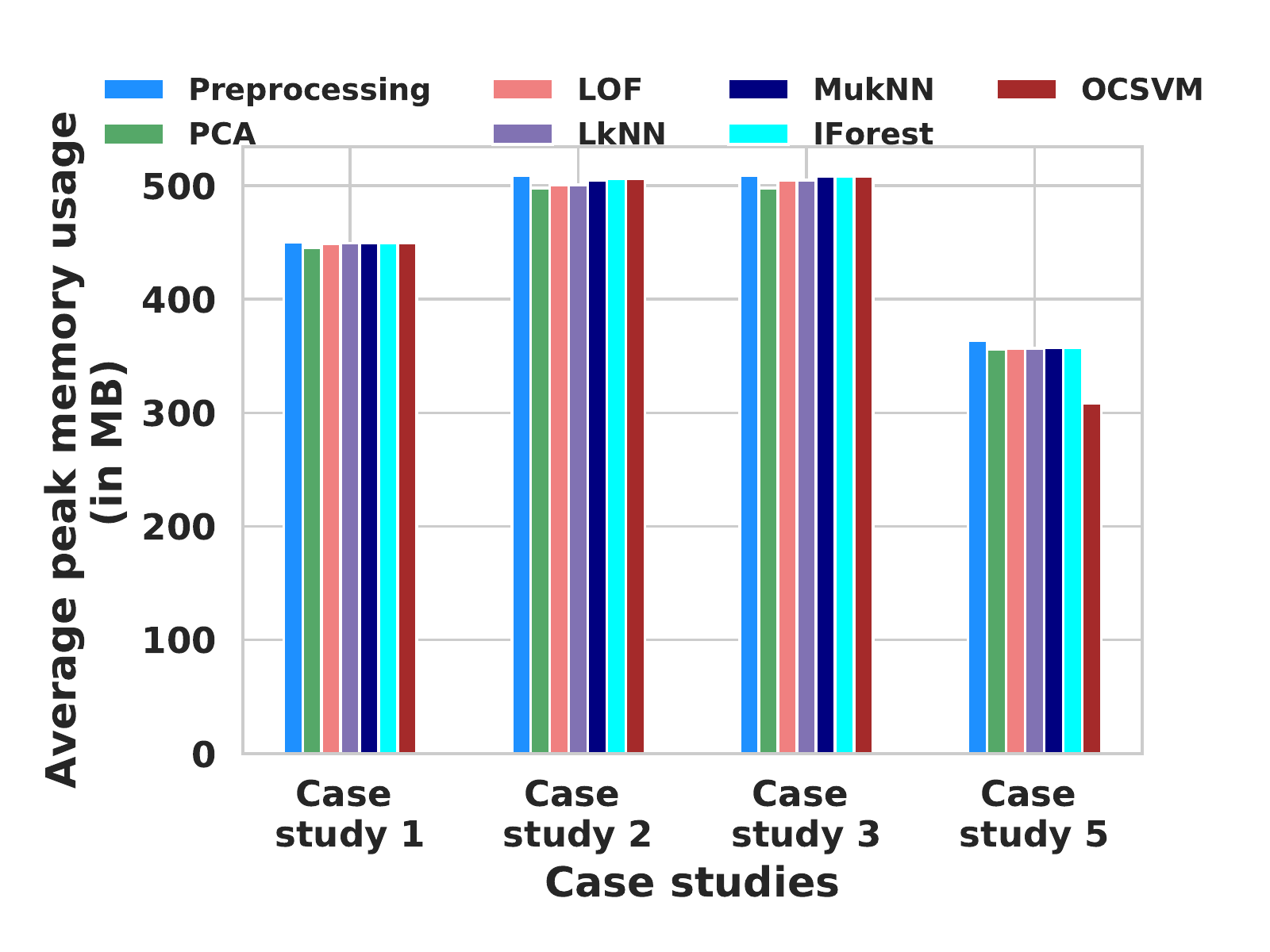}
		\caption{Peak memory usage comparison\label{fig:ad_memory}}
	\end{subfigure}
	\caption{(\subref{fig:sample_num}) shows total number of message aggregate samples for different length message sequences for different debugging case studies.~(\subref{fig:ad_runtime}) and~(\subref{fig:ad_memory}) demonstrate that our diagnosis methodology is computationally efficient in terms of runtime and peak memory usage across six different outlier detection algorithms for each of the case studies.\label{fig:basic_comparison_detection}}
	\vspace{-6mm}
\end{figure*}

\subsection{Comparison of our method to existing signal selection methods}\label{sec:comfunccov}

To demonstrate that existing Register Transfer Level signal selection methods cannot select messages in system level flows,  we compare our approach with an SRR-based method \cite{cite:basu2011efficient} and a PageRank based method \cite{cite:conf/iccad/MaPJRV15}. {\bf We could not apply existing SRR based methods on the OpenSPARC T2, since these methods are unable to scale. We use a smaller USB design for comparison with our method.} In the USB~\cite{cite:USB} design we consider a usage scenario consisting of two flows. Table~\ref{table:USBSelcetedSignals} shows that our (mutual information gain based) method selects all of {\tt token\_pid\_sel}, {\tt data\_pid\_sel} and other important interface signals for system level debugging. SigSeT, on the other hand selects signals which are not useful for system level debugging. Our messages are composed of interface signals, and achieve a flow specification coverage of {\bf 93.65\%}, whereas messages composed of interface signals selected by SigSeT and PRNet have a low flow specification coverage of {\bf 9\%} and {\bf 23.80\%} respectively.

\subsection{Selection of important messages by our method}\label{sec:selmsgbug}

For evaluation purposes, we use {\em bug coverage} as a metric, to determine which messages are important.  A message is said to be {\em affected} by a bug if its value in an execution of the buggy design differs from its value in an execution of the bug free design. Intuitively, if multiple bugs are affecting a message, it is highly likely that message is a part of multiple design paths. The {\em bug coverage} of a message is defined as the total number of bugs that affects a message, expressed as a fraction of the total number of injected bugs. From debugging perspective, a message is {\em important} if it is affected by very few bugs implying that the message symptomatizes subtle bugs. Table~\ref{table:InjBugMsgSel} confirms that post-Silicon bugs are subtle and tend to affect no more than 4 messages each. Column 4, 5 and 6 of Table~\ref{table:InjBugMsgSel} show that our method was able to select important messages from the interleaved flow to debug subtle bugs.

Table~\ref{table:InjBugMsgSel} shows that message {\em m15} is affected by four bugs and message {\em m9} is affected by two bugs, but due to their size being wider than 32 bits trace buffer, our method does not select them.

\subsection{Effectiveness of selected messages in debugging usage scenarios}\label{sec:debugcasestudy}

Every message is sourced by an IP and reaches a destination IP. Bugs are injected into specific IPs (Table~\ref{table:bugdetails}). During debug, sequences of IPs are explored from the point a bug symptom is observed, to find the buggy IP. An IP pair ($<$source IP, destination IP$>$) is {\em legal} if a message is passed between them. We use the number of legal IP pairs investigated during debug as a metric for selected messages. 
Table~\ref{table:DebuggingSteps} shows that we investigated {\bf  an average of 54.67\%} of the total legal IP pairs, implying that our selected messages help us focus on a fraction of the legal IP pairs.

To debug a buggy execution, we start with the traced message in which a bug symptom is observed and backtrack to other traced messages. The choice of which traced message to investigate is pseudo-random and guided by the participating flows.

Figure~\ref{fig:cumulative_cause_eli}(a) plots the number of such investigated traced messages and the corresponding candidate legal IP pairs that are eliminated with each traced message. Figure~\ref{fig:cumulative_cause_eli}(b) shows a similar relationship between the traced messages and the candidate root causes, \ie, the architecture level functions that might have caused the bug to manifest in the traced messages. Both graphs show that with more traced messages, more candidate legal IP pairs as well as candidate root causes are progressively eliminated. This implies that every one of our traced messages contributes to the debug process.

\begin{figure*}
	\centering
	\begin{subfigure}[b]{0.30\textwidth}
        \includegraphics[scale=0.35]{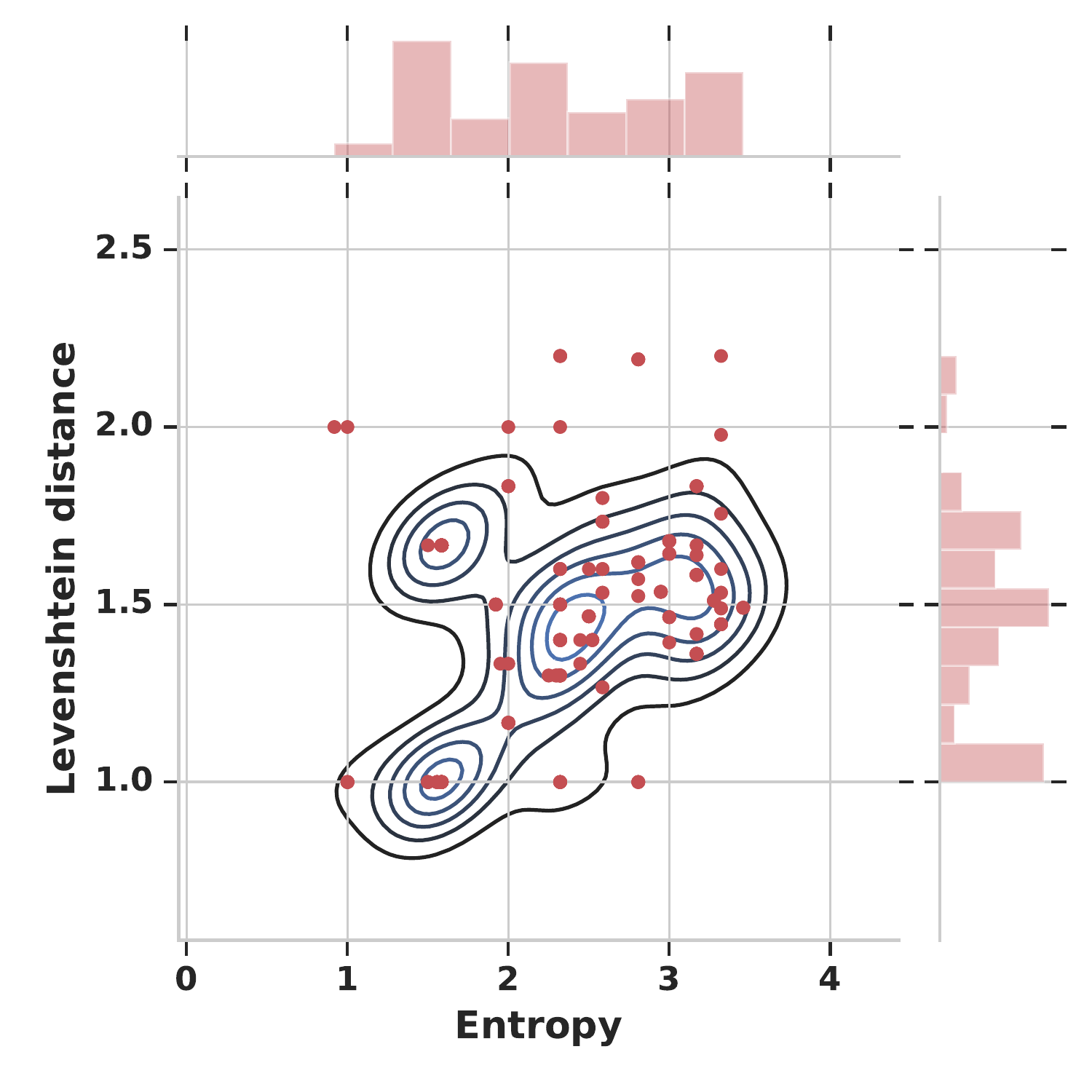}
		\caption{Case study 1 ($k = 5$)\label{fig:test1_en_ldis_joint}}
	\end{subfigure}
	\hfill
	\begin{subfigure}[b]{0.30\textwidth}
        \includegraphics[scale=0.35]{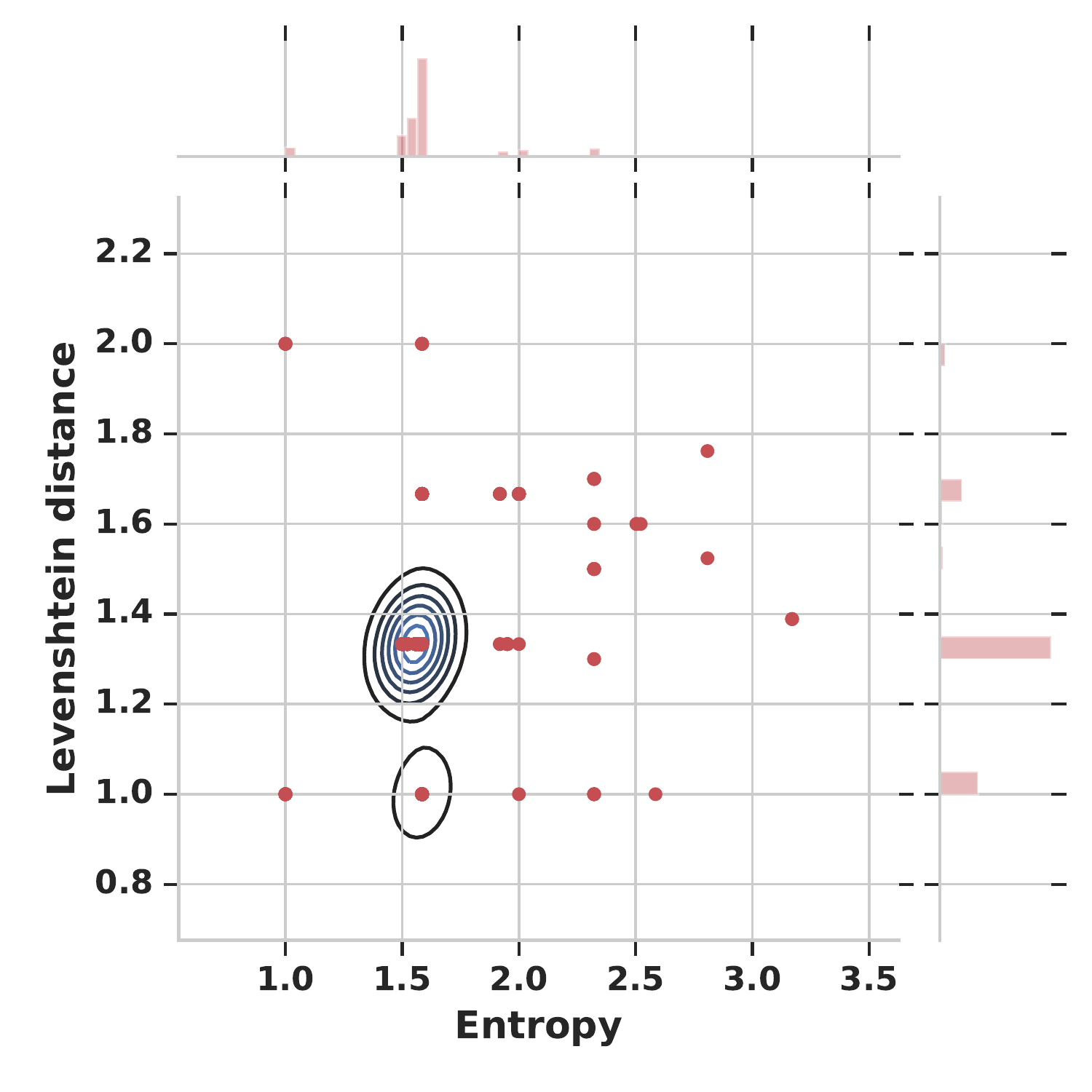}
		\caption{Case study 3 ($k = 5$)\label{fig:test3_en_ldis_joint}}
	\end{subfigure}
	\hfill
		\begin{subfigure}[b]{0.30\textwidth}
        \includegraphics[scale=0.35]{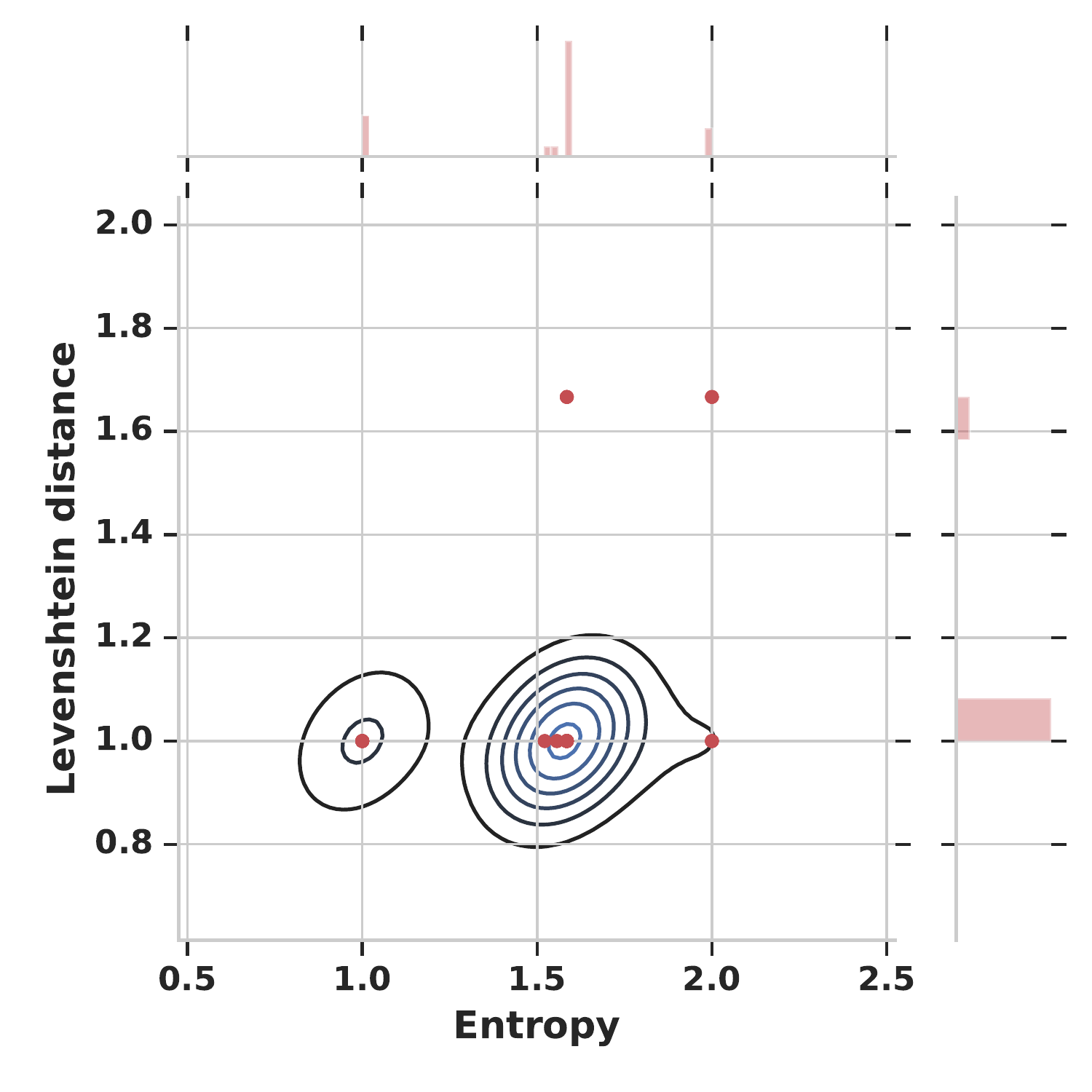}
		\caption{Case study 5 ($k = 5$)\label{fig:test5_en_ldis_joint}}
	\end{subfigure}
	\caption{(\subref{fig:test1_en_ldis_joint}), (\subref{fig:test3_en_ldis_joint}), and (\subref{fig:test5_en_ldis_joint}) show that the engineered features demarcate normal and anomalous message aggregates.\label{fig:en_ldis_joint}}
	\vspace{-5mm}
\end{figure*}

Figure~\ref{fig:CauseDistrib} shows that traced messages were able to prune out a large number of potential root causes in all five case studies. Our traced messages pruned out an {\bf average of 78.89\%} ({\bf max. 88.89\%}) of candidate root causes.


\section{Experimental results on debug and diagnosis}\label{sec:exp_results_detection}

In this section we provide insights into our bug diagnosis methodology to debug five different buggy case studies across three usage scenarios of the OpenSPARC T2 SoC. For these experiments, we have used $g = 100000$ cycles and varied $k$ from two to the number of valid IP pairs (\cf,~\Cref{table:DebuggingSteps}) for each of the case studies. The number of message aggregate samples for different lengths of message sequences for each of the outlier detection algorithm per debugging case study is shown in~\Cref{fig:sample_num}.

\subsection{Computational efforts for data preprocessing and outlier message sequence diagnosis}\label{sec:comp_effi}

In this experiment, we show scalability of the automated diagnosis methodology in terms of runtime and peak memory usage.
~\Cref{fig:ad_runtime} and~\Cref{fig:ad_memory} show runtime and peak memory usage for preprocessing and outlier detection algorithms. To calculate the average runtime and average peak memory usage of each of the outlier detection algorithms, we ran each of them {\em 20} times and calculated the average value.

Preprocessing trace message data to create message sequence aggregates incurred a runtime of {\em up to 44.3 seconds} ({\em average 10.8 seconds}) and peak memory usage of {\em up to 508.7 MB} ({\em average 457.73 MB}). To run each of the outlier detection algorithms on the processed message aggregates incurred only {\em up to 18.91 seconds} ({\em average 2.77 seconds}) and peak memory usage of {\em up to 508.2 MB} ({\em average 451.27 MB}). Since preprocessing has {\em up to 443$\times$} ({\em average 3$\times$}) more runtime than the running each of the outlier detection algorithms, we showed runtime in the $log_{10}$ scale in the~\Cref{fig:ad_runtime}.

{\bf This experiment shows that our trace-data preprocessing and diagnosis is computationally efficient.}

\subsection{Validity of entropy and Levenshtein distance as engineered feature for outlier message sequence diagnosis}\label{sec:val_ldist_entro}

\begin{figure*}
    \centering
	\begin{subfigure}[b]{0.30\textwidth}
	    \centering
        \includegraphics[scale=0.35]{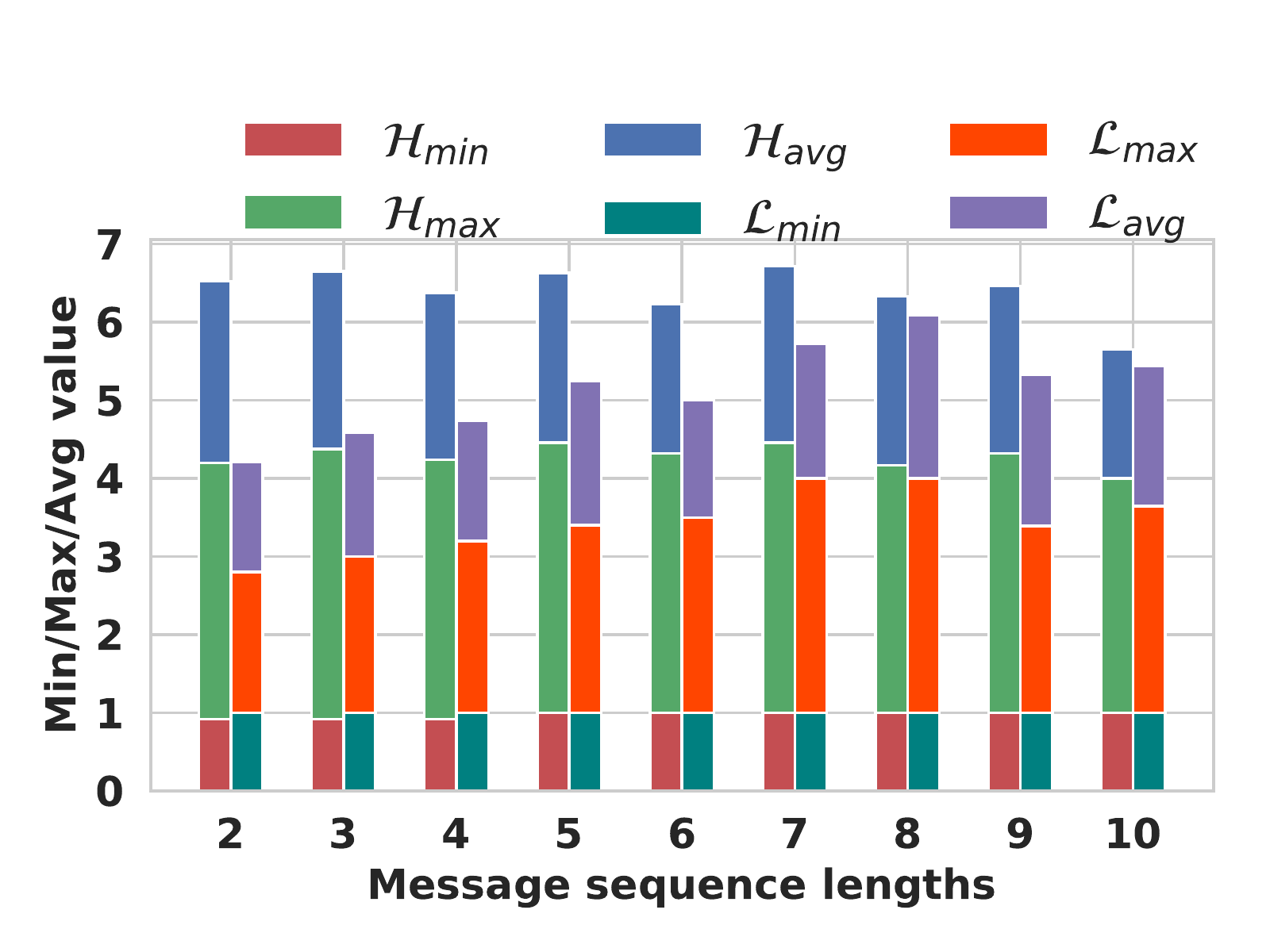}
		\caption{Case study 1\label{fig:test1_min_max_entropy}}
	\end{subfigure}
    \hspace{4mm}
	\begin{subfigure}[b]{0.30\textwidth}
	    \centering
        \includegraphics[scale=0.35]{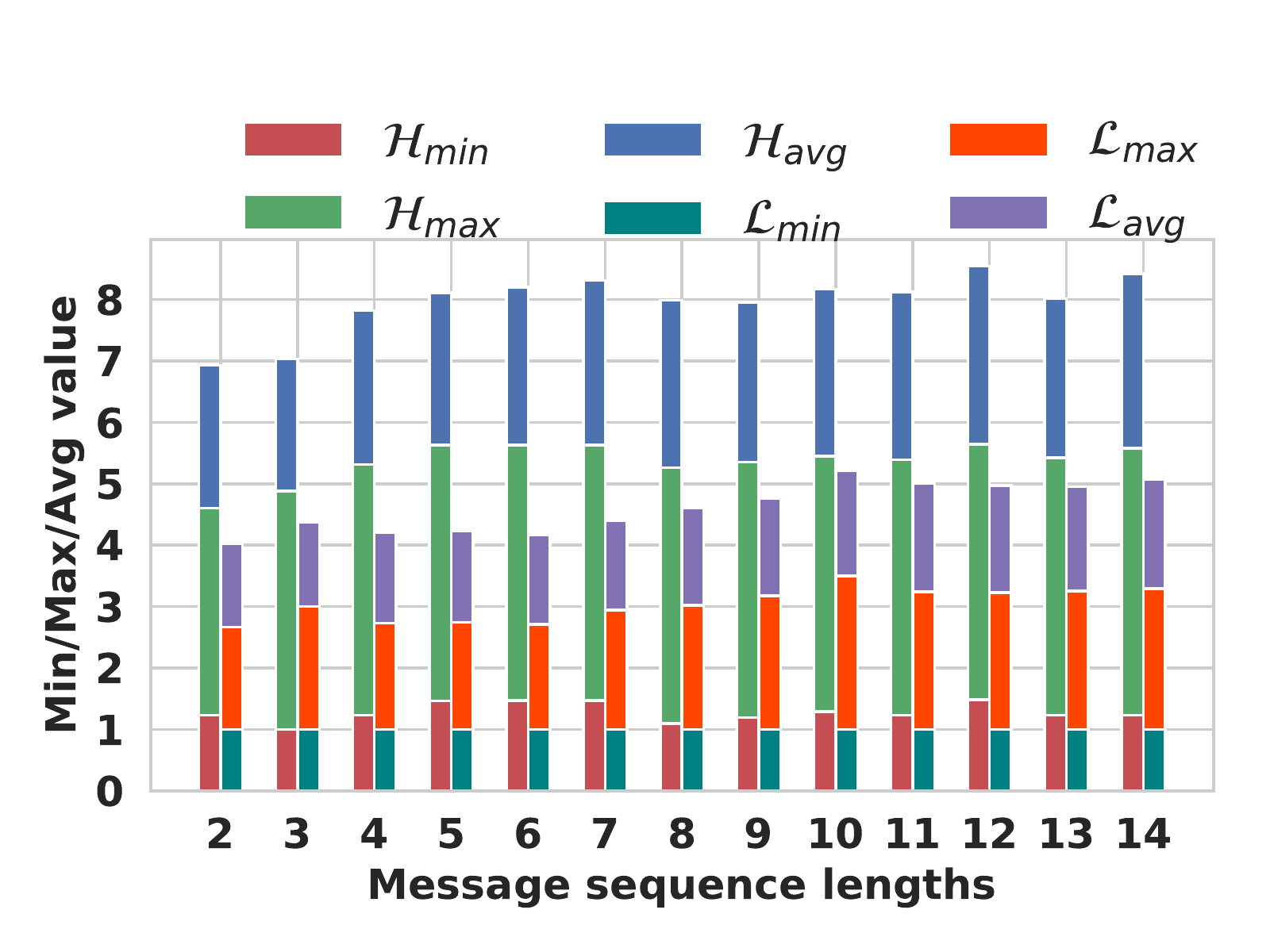}
		\caption{Case study 2\label{fig:test2_min_max_entropy}}
	\end{subfigure}
	\hspace{4mm}
	\begin{subfigure}[b]{0.30\textwidth}
	    \centering
        \includegraphics[scale=0.35]{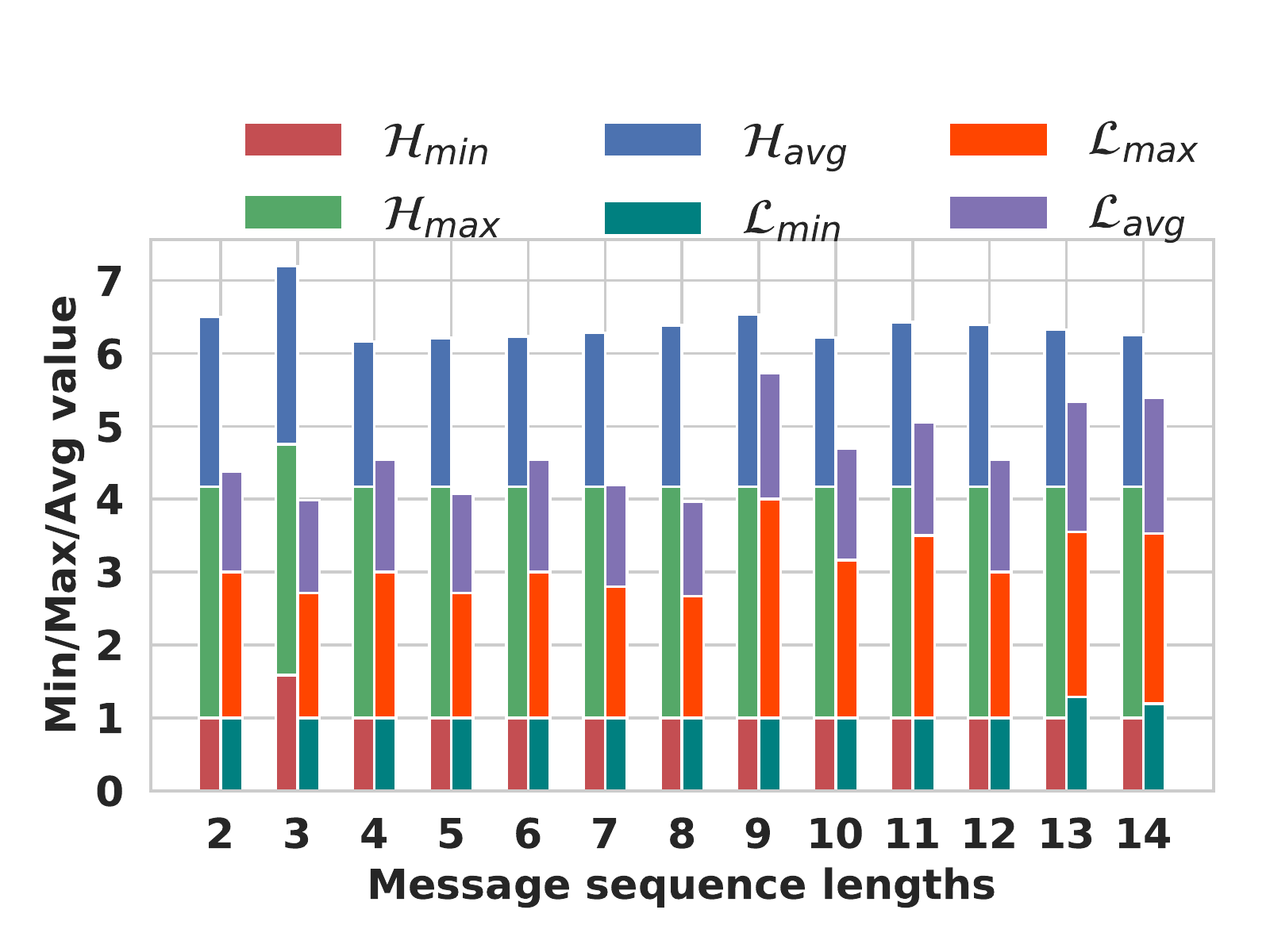}
		\caption{Case study 3\label{fig:test3_min_max_entropy}}
	\end{subfigure}
	\caption{(\subref{fig:test1_min_max_entropy}), (\subref{fig:test2_min_max_entropy}), and (\subref{fig:test3_min_max_entropy}) show that the minimum, maximum, and average value of engineered features are high for anomalous message aggregates irrespective of message sequence lengths. $\langle \mathcal{H}_{min}, \mathcal{H}_{max}, \mathcal{H}_{avg} \rangle$: Minimum, maximum, average entropy. $\langle \mathcal{L}_{min}, \mathcal{L}_{max}, \mathcal{L}_{avg} \rangle$: Minimum, maximum, average Levenshtein distance.\label{fig:min_max_entropy}}
\end{figure*}

In this experiment, we analyze the effectiveness of {\em entropy} and {\em Levenshtein distance} to identify message aggregates that contain anomalous message sequences. In~\Cref{fig:en_ldis_joint} we show joint probability distribution of entropy and Levenshtein distance and in~\Cref{fig:min_max_entropy} we show minimum, maximum, and average of entropy and Levenshtein distance of anomalous message aggregates across different length message sequences for three different debugging case studies.


As shown in~\Cref{fig:en_ldis_joint}, in the engineered feature space, message aggregates for normal behavior form a dense cluster whereas anomalous message sequences are sparsely distributed and are placed at a distance from the normal message aggregates. Further,~\Cref{fig:min_max_entropy} shows that message aggregates that contain anomalous message sequences have entropy of {\em up to 4.3482} ({\em average 2.08}) and Levenshtein distance of {\em up to 3.0} ({\em average 1.5734}).

{\bf This experiment validates that entropy and Levenshtein distance are valuable and effective engineered features in demarcating the anomalous message aggregates from normal message aggregates.}

\subsection{Agreements among different outlier detection algorithms in detecting outlier message sequences}\label{sec:agreement_outlier}

In this experiment, we assess the extent of agreement between anomalies identified by various outlier algorithms (\cf,~\Cref{sec:anomalyalgo}). Since this set of algorithms uses different methods for outlier detection, we surmise that the confidence in an anomalous message aggregate is higher, if multiple outlier detection algorithms identify it as such.
For this analysis, we consider the top 10\% of anomalous message aggregates per outlier detection algorithm per case study.



Our analysis showed that six outlier detection algorithms agree for a total of six anomalous message aggregates that diagnose {\em 13.33\%} of injected bugs, five outlier detection algorithms agree for a total of 17 anomalous message aggregates that diagnose {\em 53.33\%} of injected bugs, three outlier detection algorithms agree for a total of six anomalous message aggregates that diagnose {\em 20\%} of injected bugs, two outlier detection algorithms agree for a total of six anomalous message aggregates that diagnose {\em 26.6\%} of injected bugs.

{\bf This experiment shows that our engineered features are generic to characterize anomalies such that multiple outlier detection algorithms agree on a large number of anomalies that diagnose multiple bugs. This observation motivated us to use a comprehensive anomaly score to rank message aggregates.} We explain our comprehensive anomaly score calculation in~\Cref{sec:comprehensive}.

\begin{table*}
  \centering
  \caption{Diagnosis statistics for different outlier detection algorithms for different case studies using OpenSPARC T2 SoC~\cite{cite:SPARCT2Vol1, cite:SPARCT2Vol2}. {\bf PCA}: Principal Component Analysis~\cite{cite:pca}. {\bf LOF}: Local Outlier Factor based algorithm~\cite{cite:lof}. {\bf LkNN}: $k$-Nearest Neighbor using largest distance as metric~\cite{cite:knn1, cite:knn2}. {\bf MukNN}: k-Nearest Neighbor using mean distance as a metric~\cite{cite:knn1, cite:knn2}. {\bf OCSVM}: One-class Support Vector Machine~\cite{cite:ocsvm}. {\bf D}: Fraction of injected bugs diagnosed by an outlier detection algorithm. {\bf $t_p$}: Total number of true positive message sequences. {\bf $f_p$}: Total number of false positive message sequences (no more than 37\% anomalous message sequences). {\bf P}: Precision of an outlier detection algorithm. {\bf OS}: Overall diagnosis statistics for each of the outlier detection algorithm per debugging case study.\label{table:precisiondetail1}}
  \setlength\tabcolsep{5pt}
  \begin{tabular}{|p{0.5cm}|p{0.35cm}|p{0.35cm}|
  p{0.35cm}|p{0.35cm}|p{0.35cm}|p{0.35cm}|p{0.35cm}|p{0.35cm}|p{0.35cm}|p{0.35cm}|p{0.35cm}|
  p{0.35cm}|p{0.35cm}|p{0.35cm}|p{0.35cm}|p{0.35cm}|p{0.35cm}|p{0.35cm}|
  p{0.35cm}|p{0.35cm}|p{0.35cm}|p{0.35cm}|p{0.35cm}|p{0.35cm}|}
    \hline
    {\bf Case} & \multicolumn{4}{|c|}{\bf IForest} & \multicolumn{4}{|c|}{\bf PCA} & \multicolumn{4}{|c|}{\bf LOF} & \multicolumn{4}{|c|}{\bf LkNN} & \multicolumn{4}{|c|}{\bf MukNN} & \multicolumn{4}{|c|}{\bf OCSVM} \\
    \cline{2-25}
    {\bf study} & {\bf D} & {\bf $t_p$} & {\bf $f_p$} & {\bf P} & {\bf D} & {\bf $t_p$} & {\bf $f_p$} & {\bf P} & {\bf D} & {\bf $t_p$} & {\bf $f_p$} & {\bf P} & {\bf D} & {\bf $t_p$} & {\bf $f_p$} & {\bf P} & {\bf D} & {\bf $t_p$} & {\bf $f_p$} & {\bf P} & {\bf D} & {\bf $t_p$} & {\bf $f_p$} & {\bf P}\\
    \hline\hline
    1 & 0.75 & 9 & 4 & 0.69 & 0.25 & 7 & 3 & 0.7 & 0.5 & 2 & 2 & 0.5 & 0.25 & 18 & 4 & 0.82 & 0.75 & 9 & 4 & 0.69 & 0.75 & 20 & 6 & 0.77 \\
    \hline
    2 & 0.67 & 17 & 10 & 0.63 & 0.34 & 24 & 9 & 0.73 & 0.34 & 12 & 1 & 0.92 & 0.34 & 24 & 9 & 0.73 & 0.34 & 12 & 8 & 0.6 & 0.34 & 24 & 9 & 0.73 \\
    \hline
    3 & 0.34 & 6 & 4 & 0.6 & 0.34 & 6 & 4 & 0.6 & 0.34 & 4 & 0 & 1.0 & 0.67 & 10 & 3 & 0.77 & 0.67 & 10 & 3 & 0.77 & 0.34 & 6 & 4 & 0.6\\
    \hline
    4 & 1.0 & 7 & 3 & 0.7 & 0.34 & 6 & 4 & 0.6 & 0.34 & 3 & 2 & 0.6 & 0.34 & 9 & 3 & 0.75 & 0.67 & 9 & 3 & 0.75 & 0.67 & 8 & 2 & 0.8\\
    \hline
    5 & 1.0 & 8 & 2 & 0.8 & 1.0 & 8 & 2 & 0.8 & 1.0 & 8 & 2 & 0.8 & 1.0 & 9 & 3 & 0.75 & 1.0 & 9 & 3 & 0.75 & 1.0 & 8 & 2 &  0.8\\
    \hline\hline
    {\bf AS} & {\scriptsize 2.2} & {\scriptsize 14} & {\scriptsize 9.4} & {\scriptsize 0.69} & {\scriptsize 1.2} & {\scriptsize 14.6} & {\scriptsize 10.2} & {\scriptsize 0..69} & {\scriptsize 1.4} & {\scriptsize 7.2} & {\scriptsize 5.8} & {\scriptsize 0.76} & {\scriptsize 1.4} & {\scriptsize 18.4} & {\scriptsize 14} & {\scriptsize 0.76} & {\scriptsize 2} & {\scriptsize 14} & {\scriptsize 9.8} & {\scriptsize 0.71} & {\scriptsize 1.8} & {\scriptsize 17.8} & {\scriptsize 13.2} & {\scriptsize 0.74} \\
    \hline
  \end{tabular}
  \vspace{-2mm}
\end{table*}

\subsection{Comparison of precision of different outlier detection algorithms in detecting outlier message sequences}\label{sec:precision_od}

In this experiment, we compare the {\em precision} (\cf,~\Cref{def:precision}), {\em recall} (\cf,~\Cref{def:recall}), and {\em accuracy} (\cf,~\Cref{def:accuracy}) of each of the outlier detection algorithms in diagnosing anomalous messages sequences per debugging case study. In~\Cref{table:precisiondetail1}, we show the fraction of injected  bugs diagnosed, and the number of true positive and false positive candidate anomalous message sequences identified for each of the outlier detection algorithm per debugging case study. In~\Cref{table:precisiondetail2}, we show the fraction of total number of injected bugs diagnosed, total number of true positive, false positive, true negative, and false negative candidate anomalous message sequences identified across all of the outlier detection algorithms per debugging case study. For this analysis, we considered only the top 10\% anomalous message aggregates identified by each of the outlier detection algorithm per debugging case study.

Our analysis shows that IForest, MukNN, and OCSVM consistently performed better in anomalous message sequence diagnosis as compared to the other three algorithms PCA, LOF, and LkNN. Each of the outlier detection algorithm diagnosed {\em up to 100\%} of injected bugs. IForest diagnosed on an {\em average 73\%} of injected bugs with a precision of {\em up to 0.8} ({\em average 0.69}), MukNN diagnosed on an {\em average 67\%} of injected bugs with a precision of {\em up to 0.77} ({\em average 0.70}), and OCSVM diagnosed on an {\em average 67\%} of injected bugs with a precision of {\em up to 0.8} ({\em average 0.74}) per debugging case study. On the other hand, PCA diagnosed on an average {\em average 40\%} of injected bugs with a precision of {\em up to 0.8} ({\em average 0.69}), LOF diagnosed on an {\em average 47\%} of injected bugs with a precision of {\em up to 1.0} ({\em average 0.81}), and LkNN diagnosed on an {\em average 47\%} of injected bugs with a precision of {\em up to 0.82} ({\em average 0.76}) per debugging case study. Further analysis shows (\cf,~\Cref{table:precisiondetail2}) our automated diagnosis technique was able to detect {\em up to 100\%} ({\em average 81.8\%}) of injected bugs with a precision of {\em up to 0.769} ({\em average 0.756}) per debugging case study.

In~\Cref{table:precisiondetail2}, we also show the recall and the accuracy metric per debugging case study. Our diagnosis methodology achieved {\em up to 0.69} ({\em average 0.46}) recall and {\em up to 0.56} ({\em average 0.39}) accuracy. We note that in~\Cref{table:precisiondetail2} the value of recall and accuracy are relatively small. This is due to the fact that we are only considering the top 10\% anomalous message aggregates for this analysis. Consequently, the $t_p$ in the numerator is calculated from those top 10\% anomalous message aggregates whereas $f_n$ and $t_n$ are calculated based on the entire set of message aggregates. Consequently, the numerators are much smaller than the denominators (\cf,~\Cref{def:recall} and~\Cref{def:accuracy}) which results in a small value of recall and accuracy.

{\bf This experiment shows that our automated diagnosis methodology using engineered features is effective in identifying complex and subtle bugs with high precision.}


\begin{table}
  \centering
  \caption{Overall statistics of automated debugging across all outlier detection algorithms across all case studies. {\bf D}: Fraction of injected bugs detected. {\bf P}: Precision. {\bf R}: Recall. {\bf A}: Accuracy.\label{table:precisiondetail2}}
    \begin{tabular}{|c|c|c|c|c|c|c|c|c|}
    \hline
    {\bf Case} & \multirow{2}{*}{\bf D} & \multicolumn{4}{|c|}{\bf Sequences} & \multirow{2}{*}{\bf P} & \multirow{2}{*}{\bf R} & \multirow{2}{*}{\bf A}\\
    \cline{3-6}
    {\bf Study} & {} & {\bf $t_p$} & {$t_n$} & {$f_p$} & {$f_n$} & {} & {} & {} \\
    \hline\hline
    1 & 0.75 & 20 & 2 & 6 & 54 & 0.769 & 0.27 & 0.25 \\
    \hline
    2 & 0.67 & 29 & 2 & 11 & 24 & 0.725 & 0.54 & 0.45 \\
    \hline
    3 & 0.67 & 10 & 2 & 3 & 22 & 0.769 & 0.32 & 0.28 \\
    \hline
    4 & 1.0 & 20 & 0 & 6 & 22 & 0.769 & 0.48 & 0.42 \\
    \hline
    5 & 1.0 & 9 & 1 & 3 & 4 & 0.75 & 0.69 & 0.56 \\
    \hline
  \end{tabular}
\end{table}

\subsection{Improvement in diagnosis over manual debugging}\label{sec:improvement}

In this experiment, we analyze the improvement in diagnosis in terms of number of injected bugs diagnosed and diagnosis time over manual debugging.~\Cref{table:improvement_over_manual_debugging} (column 7 and column 8) summarizes the diagnosis improvement. We were able to diagnose {\em up to 66.7\% more injected bugs} ({\em average 46.67\%}) with {\em up to 847$\times$} ({\em average 464.35$\times$}) less diagnosis time.

{\bf This experiment shows that our automated bug diagnosis is effective and expedites debugging.}

\subsection{Comprehensive ranking of outlier message sequences}\label{sec:comprehensive}

In~\Cref{sec:precision_od}, our experimental results showed that IForest, OCSVM, and MukNN are the three most effective outlier detection algorithms among six for diagnosing useful anomalous message sequences that can help in debugging. Each of the IForest, OCSVM, and MukNN (\cf,~\Cref{sec:anomalyalgo}) detect anomalous message aggregates based on a different perspective. IForest selects an anomalous message aggregate based on {\em shorter path lengths} created by {\em random selection of a feature} and {\em recursive partitioning of the feature data}. OCSVM selects an anomalous message aggregate by solving an optimization problem to find a {\em maximal margin hyperplane} that best separates anomalous message aggregates. MukNN (\ie, $k$-NN with mean distance as metric) selects an anomalous message aggregate based on a aggregate's local density and the distance to its $k^{th}$ nearest neighbor.

Consequently, to incorporate these different perspectives into our diagnosis methodology, we use a heuristic combination of outlier scores from each of the above three algorithms for each of the message aggregate. We found that a linear combination of outlier scores of a message aggregate is in closer agreement with our empirical findings than relying on outlier score of a message aggregate from each of the individual algorithms. Let $x$ be a message aggregate, $Ano(x)$ be the comprehensive outlier score of $x$, and $IForest(x)$, $OCSVM(x)$, and $MukNN(x)$ be the outlier score of $x$ using the IForest, OCSVM, and MukNN algorithm respectively. We define $Ano(x)$ as $Ano(x)  = (IForest(x) + OCSVM(x) + MukNN(x))/3$. In our experiments, we rank anomalous message aggregates based on the comprehensive outlier score defined above.




\begin{table}
  \centering
  \caption{Summary of detection improvements achieved using automated detection technique over manual debugging. {\bf N}: Number of symptomatic message sequences identified. {\bf T}: Time taken to identify a symptomatic message sequence. {\bf D}: Improvement in terms of number of additional detected bugs as a fraction of injected bugs. {\bf t} : Improvement in detection time. {$\oslash$}: Not available.\label{table:improvement_over_manual_debugging}}
  \begin{tabular}{|>{\centering}m{0.5cm}|>{\centering}m{0.5cm}|>{\centering}m{0.8cm}|
  >{\centering}m{0.5cm}|>{\centering}m{0.5cm}|>{\centering}m{0.8cm}|
  >{\centering}m{0.8cm}|p{0.8cm}|}
    \hline
    {\bf Case} & {\bf Bug} & \multicolumn{2}{|c|}{\bf Manual} & \multicolumn{2}{|c|}{\bf Automated} & \multicolumn{2}{|c|}{\bf Improvement}\\
    \cline{3-8}
    {\bf study} & {\bf ID} & {\bf N} & {\bf T} & {\bf N} & {\bf T} & {\bf D} & {\bf t} \\
    & & & {\bf (Hrs)} & & {\bf (Secs)} & & \\
    \hline\hline
     \multirow{4}{*}{1} & 1 & 1 & 8 & 18 & \multirow{3}{*}{61.4} & \multirow{4}{*}{50\%}
                        & \multirow{4}{*}{469.1$\times$} \\
                        \cline{2-5}
                        & 28 &  \multicolumn{1}{|c|}{$\oslash$} & \multicolumn{1}{|c|}{$\oslash$} & \multirow{2}{*}{2} & & & \\
                        \cline{2-4}
                        & 29 & \multicolumn{1}{|c|}{$\oslash$} &  \multicolumn{1}{|c|}{$\oslash$} &  & & & \\
                        \cline{2-6}
                        & 36 &  \multicolumn{1}{|c|}{$\oslash$} & \multicolumn{1}{|c|}{$\oslash$} & \multicolumn{1}{|c|}{$\oslash$} & \multicolumn{1}{|c|}{$\oslash$} & & \\
    \hline
     \multirow{3}{*}{2} & 17 & \multicolumn{1}{|c|}{$\oslash$} &
                        \multicolumn{1}{|c|}{$\oslash$} & 5 & \multirow{2}{*}{58.5} & \multirow{3}{*}{33.3\%} & \multirow{3}{*}{184.61$\times$}\\
                        \cline{2-5}
                        & 18 & 1 & 3 & 24 & & & \\
                        \cline{2-6}
                        & 25 & \multicolumn{1}{|c|}{$\oslash$} & \multicolumn{1}{|c|}{$\oslash$} & \multicolumn{1}{|c|}{$\oslash$} & \multicolumn{1}{|c|}{$\oslash$} & & \\
    \hline
     \multirow{3}{*}{3} & 5 &
                        \multicolumn{1}{|c|}{$\oslash$} & \multicolumn{1}{|c|}{$\oslash$} & \multicolumn{1}{|c|}{$\oslash$} & \multicolumn{1}{|c|}{$\oslash$} & \multirow{4}{*}{33.3\%} & \multirow{3}{*}{847.05$\times$} \\
                        \cline{2-6}
                        & 8 & 1 & 14 & 6 & \multirow{2}{*}{59.5} & & \\
                        \cline{2-5}
                        & 37 & \multicolumn{1}{|c|}{$\oslash$} & \multicolumn{1}{|c|}{$\oslash$} & 4 & & & \\
    \hline
     \multirow{3}{*}{4} & 5  & 1 & 6 & 14 & \multirow{3}{*}{57.5} &
                        \multirow{3}{*}{66.7\%} & \multirow{3}{*}{375.65$\times$}\\
                        \cline{2-5}
                        & 8  & \multicolumn{1}{|c|}{$\oslash$} & \multicolumn{1}{|c|}{$\oslash$} & 3 & & & \\
                        \cline{2-5}
                        & 37 & \multicolumn{1}{|c|}{$\oslash$} & \multicolumn{1}{|c|}{$\oslash$} & 3 & & & \\
    \hline
     \multirow{2}{*}{5} & 24  & \multicolumn{1}{|c|}{$\oslash$} &
                        \multicolumn{1}{|c|}{$\oslash$} & 3 & \multirow{2}{*}{48.5} & \multirow{2}{*}{50\%} & \multirow{2}{*}{445.36$\times$}\\
                        \cline{2-5}
                        & 39  & 1 & 6 & 6  & & & \\
    \hline
  \end{tabular}
\end{table}

\begin{table*}
  \centering
  \scriptsize
  \caption{Representative potential root causes for one case study. Rest of the root causes are omitted due to lack of space. Remaining case studies are available in~\cite{cite:csltechreport}\label{table:debug_reasons_scene_1}}
  \begin{tabular}{|>{\centering}m{2.0cm}|>{\centering}m{8.5cm}|p{6.5cm}|}
   \hline
    {\bf Selected Messages} & {\bf Potential Causes} & {\bf Potential Implication} \\
    \hline
     {\tt reqtot,grant, mondoacknack, siincu, piowcrd} & {\bf 1.} Mondo request forwarded from DMU to SIU's bypass queue instead of ordered queue & {\bf 1.}  Mondo interrupt not serviced \\
     \cline{2-3}
     {\tt dmusiidata. cputhreadid} & {\bf 2.} Invalid Mondo payload forwarded to NCU from DMU via SIU & {\bf 2.} Interrupt assigned to wrong CPU ID and Thread ID \\
     \cline{2-3}
    \hline
   \end{tabular}
   \vspace{-4mm}
\end{table*}

\section{Qualitative case study on effectiveness of our message selection and diagnosis methodology}\label{sec:qual_case_studies}

It is illuminating to understand a case study to appreciate the effectiveness of the selected messages and our bug detection methodology in the debugging process.

\smallskip


\noindent {\bf \em Symptom:} In this experiment we used traced messages from Table~\ref{table:debug_reasons_scene_1}. The simulation failed with an error message {\em FAIL: Bad Trap}.

\smallskip

\noindent {\bf \em Manual debug with selected messages:} We consider bug symptom causes of Table~\ref{table:debug_reasons_scene_1} to debug this case. From the observed trace messages, {\tt siincu} and {\tt piowcrd}, we identify NCU got back correct credit ID at the end of the PIO read and PIO write operation respectively. This rules out two causes out of 9. However, we cannot rule out causes related to PIO payload since a wrong payload may cause computing thread to catch {\em BAD Trap} by requesting operand from wrong memory location. Absence of trace messages {\tt mondoacknack} and {\tt reqtot} implies that NCU did not service any Mondo interrupt request and SIU did not request a Mondo payload transfer to NCU respectively. Further, there is no message corresponding to {\tt dmusiidata.cputhreadid} in the trace file, implying that DMU was never able to generate a Mondo interrupt request for NCU to process. This rules out all causes except cause {\bf 3} ({\bf 1 cause out of 9, pruning of 88.89\% of possible causes}) to explore further to find the root cause.

\smallskip

\noindent {\bf \em Manual root causing:} From~\cite{cite:SPARCT2Vol1, cite:SPARCT2Vol2}, we note that an interrupt is generated only when DMU has credit and all previous DMA reads are done. We found no prior DMA read messages and DMU had all its credit available. Absence of {\tt dmusiidata} message correct CPUID and ThreadID implies that DMU never generated a Mondo interrupt request. This makes DMU a plausible location of the root cause of the bug.


\smallskip

%
%

\smallskip

\noindent {\bf Debug with bug diagnosis methodology:} We apply our bug diagnosis methodology on the same set of trace messages as before. The methodology identified {\em five} anomalous message aggregates containing a total of 26 unique message sequences. We found 20 {\em true positive} anomalous message sequences that are symptomatic of different bugs that we injected in the design. Among these 20 anomalous message sequences, 18 message sequences were symptomatic of the bug that we identified manually. The remaining two message sequences were symptomatic of the other two injected bugs.

Clearly, while debugging manually, we were unable to detect the later two bugs because i) they were more subtle and ii) the symptomatic message sequences were extremely infrequent. Interestingly, {\em the manual debug took approximately eight hours} to diagnose one symptomatic message sequence. In comparison, the automated bug diagnosis methodology took only {\em approximately 62 seconds} (an improvement of {\em 469$\times$}) to pre-process the trace messages and to diagnose candidate anomalous message sequences using different outlier detection algorithms. Additionally, the diagnosis method was able to diagnose candidate anomalous message sequences for two more bugs, an improvement of {\em 50\%} over manual debugging (\cf,~\Cref{table:improvement_over_manual_debugging}).

{\bf This case study shows that our bug diagnosis methodology automates and expedites tedious and error-prone manual debugging process of post-silicon failures.}

%

\section{Discussions and Conclusion}\label{sec:conclusion}

In light of our experimental findings, we believe that a synergistic application of feature engineering and anomaly detection is a powerful tool for application-level post-silicon debug and diagnosis. Although the two features presented in this work capture a wide range of bugs, we concur that these sets of features are not complete and may fail to capture certain application-level bugs. Since our proposed bug diagnosis framework is very generic, one may engineer additional features to plug-in to diagnose a wider set of bugs.

In conclusion, we have presented an automated post-silicon bug diagnosis methodology for SoC use-case failures. Our solution uses the power of machine learning and feature engineering to automatically learn the buggy design behavior and the normal design behavior from the trace data by analyzing intrinsic data feature without requiring prior knowledge of the design. Our proposed diagnosis solution is highly effective and can diagnose many more bugs at a fraction of time with high precision as compared to manual debugging. We demonstrate the effectiveness of our proposed diagnosis solution using real-world debugging case studies on the OpenSPARC T2 SoC.


\begin{thebibliography}{10}

\bibitem{cite:conf/dac/AbarbanelSV14}
Y.~Abarbanel, E.~Singerman, and M.~Y. Vardi.
\newblock Validation of soc firmware-hardware flows: Challenges and solution
  directions.
\newblock In {\em The 51st Annual {DAC} '14, San Francisco, CA, USA, June 1-5,
  2014}, pages 2:1--2:4, 2014.

\bibitem{cite:ocsvm2}
M.~Amer, M.~Goldstein, and S.~Abdennadher.
\newblock Enhancing one-class support vector machines for unsupervised anomaly
  detection.
\newblock In {\em Proceedings of the ACM SIGKDD Workshop on Outlier Detection
  and Description}, pages 8--15. ACM, 2013.

\bibitem{cite:knn1}
F.~Angiulli and C.~Pizzuti.
\newblock Fast outlier detection in high dimensional spaces.
\newblock In {\em Proceedings of the 6th European Conference on Principles of
  Data Mining and Knowledge Discovery}, PKDD '02, pages 15--26, London, UK, UK,
  2002. Springer-Verlag.

\bibitem{cite:basu2011efficient}
K.~Basu and P.~Mishra.
\newblock Efficient trace signal selection for post silicon validation and
  debug.
\newblock In {\em VLSI Design (VLSI Design), 2011 24th International Conference
  on}, pages 352--357. IEEE, 2011.

\bibitem{bengio2014representation}
Y.~Bengio, A.~Courville, and P.~Vincent.
\newblock Representation learning: A review and new perspectives.
\newblock {\em arXiv preprint arXiv:1206.5538}, 2014.

\bibitem{cite:lof}
M.~M. Breunig, H.-P. Kriegel, R.~T. Ng, and J.~Sander.
\newblock Lof: Identifying density-based local outliers.
\newblock In {\em Proceedings of the 2000 ACM SIGMOD International Conference
  on Management of Data}, SIGMOD '00, pages 93--104, New York, NY, USA, 2000.
  ACM.

\bibitem{cite:od_survey2}
V.~Chandola, A.~Banerjee, and V.~Kumar.
\newblock Anomaly detection: A survey.
\newblock {\em ACM computing surveys (CSUR)}, 41(3):15, 2009.

\bibitem{cite:chatterjee2011simulation}
D.~Chatterjee, C.~McCarter, and V.~Bertacco.
\newblock Simulation-based signal selection for state restoration in silicon
  debug.
\newblock In {\em Computer-Aided Design (ICCAD), 2011 IEEE/ACM International
  Conference on}, pages 595--601. IEEE, 2011.

\bibitem{cite:conf/memocode/FraerKKNPSTVY14}
R.~Fraer, D.~Keren, Z.~Khasidashvili, A.~Novakovsky, A.~Puder, E.~Singerman,
  E.~Talmor, M.~Y. Vardi, and J.~Yang.
\newblock From visual to logical formalisms for soc validation.
\newblock In {\em Twelfth {ACM/IEEE} {MEMOCODE} 2014, Lausanne, Switzerland,
  October 19-21, 2014}, pages 165--174, 2014.

\bibitem{cite:od_survey1}
M.~Goldstein and S.~Uchida.
\newblock A comparative evaluation of unsupervised anomaly detection algorithms
  for multivariate data.
\newblock {\em PloS one}, 11(4):e0152173, 2016.

\bibitem{cite:hamming}
R.~W. {Hamming}.
\newblock Error detecting and error correcting codes.
\newblock {\em The Bell System Technical Journal}, 29(2):147--160, April 1950.

\bibitem{Heaton_2016}
J.~Heaton.
\newblock An empirical analysis of feature engineering for predictive modeling.
\newblock {\em SoutheastCon 2016}, 2016.

\bibitem{cite:ldistance}
{Levenshtein distance}.
\newblock \url{https://en.wikipedia.org/wiki/Levenshtein_distance}.

\bibitem{lin2014effective}
D.~Lin, T.~Hong, Y.~Li, S.~Kumar, F.~Fallah, N.~Hakim, D.~Gardner, S.~Mitra,
  et~al.
\newblock Effective post-silicon validation of system-on-chips using quick
  error detection.
\newblock {\em Computer-Aided Design of Integrated Circuits and Systems, IEEE
  Transactions on}, 33(10):1573--1590, 2014.

\bibitem{cite:pca}
M.~ling Shyu, S.~ching Chen, K.~Sarinnapakorn, and L.~Chang.
\newblock A novel anomaly detection scheme based on principal component
  classifier.
\newblock In {\em in Proceedings of the IEEE Foundations and New Directions of
  Data Mining Workshop, in conjunction with the Third IEEE International
  Conference on Data Mining (ICDM’03}, pages 172--179, 2003.

\bibitem{cite:iforest2}
F.~T. Liu, K.~M. Ting, and Z.-H. Zhou.
\newblock Isolation forest.
\newblock In {\em Proceedings of the 2008 Eighth IEEE International Conference
  on Data Mining}, ICDM '08, pages 413--422, Washington, DC, USA, 2008. IEEE
  Computer Society.

\bibitem{cite:iforest1}
F.~T. Liu, K.~M. Ting, and Z.-H. Zhou.
\newblock Isolation-based anomaly detection.
\newblock {\em ACM Trans. Knowl. Discov. Data}, 6(1):3:1--3:39, Mar. 2012.

\bibitem{cite:conf/iccad/MaPJRV15}
S.~Ma, D.~Pal, R.~Jiang, S.~Ray, and S.~Vasudevan.
\newblock Can't see the forest for the trees: State restoration's limitations
  in post-silicon trace signal selection.
\newblock In {\em Proceedings of {ICCAD} 2015, Austin, TX, USA, November 2-6,
  2015}, pages 1--8, 2015.

\bibitem{Mitchell:1997:ML:541177}
T.~M. Mitchell.
\newblock {\em Machine learning}.
\newblock McGraw-Hill, Inc., New York, NY, USA, 1 edition, 1997.

\bibitem{cite:Mitra:2010:PVO:1837274.1837280}
S.~Mitra, S.~A. Seshia, and N.~Nicolici.
\newblock Post-silicon validation opportunities, challenges and recent
  advances.
\newblock In {\em Proceedings of the 47th Design Automation Conference}, DAC
  '10, pages 12--17, New York, NY, USA, 2010. ACM.

\bibitem{DBLP:books/lib/Murphy12}
K.~P. Murphy.
\newblock {\em Machine learning - {A} probabilistic perspective}.
\newblock Adaptive computation and machine learning series. {MIT} Press, 2012.

\bibitem{DBLP:conf/dac/PalSRPV18}
D.~Pal, A.~Sharma, S.~Ray, F.~M. de~Paula, and S.~Vasudevan.
\newblock Application level hardware tracing for scaling post-silicon debug.
\newblock In {\em Proceedings of the 55th Annual Design Automation Conference,
  {DAC} 2018, San Francisco, CA, USA, June 24-29, 2018}, pages 92:1--92:6,
  2018.

\bibitem{cite:patra}
P.~Patra.
\newblock {On the Cusp of a Validation Wall}.
\newblock {\em IEEE Design and. Test of Computers}, 24(2):193--196, 2007.

\bibitem{cite:journals/tvlsi/RahmaniRM17}
K.~Rahmani, S.~Ray, and P.~Mishra.
\newblock Postsilicon trace signal selection using machine learning techniques.
\newblock {\em {IEEE} Trans. {VLSI} Syst.}, 25(2):570--580, 2017.

\bibitem{cite:knn2}
S.~Ramaswamy, R.~Rastogi, and K.~Shim.
\newblock Efficient algorithms for mining outliers from large data sets.
\newblock In {\em Proceedings of the 2000 ACM SIGMOD International Conference
  on Management of Data}, SIGMOD '00, pages 427--438, New York, NY, USA, 2000.
  ACM.

\bibitem{cite:ocsvm}
B.~Sch\"{o}lkopf, J.~C. Platt, J.~C. Shawe-Taylor, A.~J. Smola, and R.~C.
  Williamson.
\newblock Estimating the support of a high-dimensional distribution.
\newblock {\em Neural Comput.}, 13(7):1443--1471, July 2001.

\bibitem{doi:10.1002/j.1538-7305.1948.tb01338.x}
C.~E. Shannon.
\newblock A mathematical theory of communication.
\newblock {\em Bell System Technical Journal}, 27(3):379--423, 1948.

\bibitem{cite:conf/dac/SingermanAB11}
E.~Singerman, Y.~Abarbanel, and S.~Baartmans.
\newblock Transaction based pre-to-post silicon validation.
\newblock In {\em Proceedings of the 48th {DAC} 2011, San Diego, California,
  USA, June 5-10, 2011}, pages 564--568, 2011.

\bibitem{cite:SPARCT2Vol1}
{OpenSPARC T2 MicroArchitecture Specification Vol 1}.
\newblock
  \url{http://www.oracle.com/technetwork/systems/opensparc/t2-07-opensparct2-socmicroarchvol1-1537750.html}.

\bibitem{cite:SPARCT2Vol2}
{OpenSPARC T2 MicroArchitecture Specification Vol 2}.
\newblock
  \url{http://www.oracle.com/technetwork/systems/opensparc/t2-08-opensparct2-socmicroarchvol2-1537751.html}.

\bibitem{cite:conf/fmcad/TalupurRE15}
M.~Talupur, S.~Ray, and J.~Erickson.
\newblock Transaction flows and executable models: Formalization and analysis
  of message passing protocols.
\newblock In {\em {FMCAD} 2015, Austin, Texas, USA, September 27-30, 2015.},
  pages 168--175, 2015.

\bibitem{cite:csltechreport}
{Zoom Out and See Better: Scalable Message Tracing for Post-Silicon SoC Debug},
  2017.
\newblock http://hdl.handle.net/2142/98857.

\bibitem{cite:USB}
{USB 2.0}, 2008.
\newblock http://opencores.org/project,usb.

\bibitem{cite:yerramili}
S.~Yerramilli.
\newblock {Addressing Post-Silicon Validation Challenge: Leverage Validation
  and Test Synergy}.
\newblock In {\em Keynote, Intl. Test Conf.}, 2006.

\bibitem{cite:zhao2019pyod}
Y.~Zhao, Z.~Nasrullah, and Z.~Li.
\newblock Pyod: A python toolbox for scalable outlier detection.
\newblock {\em arXiv preprint arXiv:1901.01588}, 2019.

\end{thebibliography}

\end{document}